\documentclass{jfm}
\usepackage{graphicx}
\usepackage{epstopdf, epsfig}
\usepackage{subcaption}
\usepackage{bm}

\shorttitle{Wavepacket Modelling of BBSAN } 
\shortauthor{M. H. Wong et al} 

\title{Wavepacket Modelling of Broadband Shock-Associated Noise in Supersonic Jets}

\author
 {
 Marcus H. Wong\aff{1}
  \corresp{\email{marcus.wong@monash.edu}},
  Peter Jordan \aff{2},
  Igor A. Maia \aff{2},
  Andr{\'e} V. G. Cavalieri \aff{3},
  Rhiannon Kirby \aff{1},
  Thales C. L. Fava \aff{4}
  \and 
  Daniel Edgington-Mitchell \aff{1}
  }

\affiliation
{
\aff{1}
Laboratory for Turbulence Research in Aerospace and Combustion, Department of Mechanical and Aerospace Engineering, Monash University, Melbourne, VIC 3800, Australia
\aff{2}
Departement Fluides, Thermique, Combustion, Institut PPRIME,
CNRS - Universit{\'e} de Poitiers - ENSMA, 86036 Poitiers, France
\aff{3}
Divis\~ao de Engenharia  Aeron{\'a}utica, Instituto Tecnol{\'o}gico de Aeron{\'a}utica, Instituto Tecnol{\'o}gico de Aeron{\'a}utica, S\~ao Jos{\'e} dos Campos, SP, 12228-900, Brazil
\aff{4}
KTH Royal Institute of Technology, Linn{\'e} FLOW Centre, Department of Mechanics, SE-10044, Stockholm, Sweden
}

\begin{document}

\maketitle

\begin{abstract}
We present a two-point model to investigate the underlying source mechanisms for broadband shock-associated noise (BBSAN) in shock-containing supersonic jets. In the model presented, the generation of BBSAN is assumed to arise from the non-linear interaction between downstream-propagating coherent structures with the quasi-periodic shock cells in the jet plume. The turbulent perturbations are represented as axially-extended wavepackets and the shock cells are modelled as a set of stationary waveguide modes. Unlike previous BBSAN models, the physical parameters describing the hydrodynamic components are not scaled using the acoustic field. Instead, the characteristics of both the turbulent and shock components are educed from large-eddy simulation and particle image velocimetry datasets. Apart from using extracted data, a reduced-order description of the wavepacket structure is obtained using parabolised stability equations (PSE). The validity of the model is tested by comparing far-field sound pressure level predictions to azimuthally-decomposed experimental acoustic data from a cold Mach 1.5 underexpanded jet. At polar angles and frequencies where BBSAN dominates, good agreement in spectral shape and sound amplitude is observed for the first three azimuthal modes. Encouraging comparisons of the radiated noise spectra, in both frequency and amplitude, reinforce the suitability of using reduced-order linear wavepacket sources for predicting BBSAN peaks. On the other hand, the mismatch in sound amplitude at inter-peak frequencies reveals the role of wavepacket jitter in the underlying sound generating mechanism. 

\end{abstract}


\section{Introduction}
The intense noise radiated by high-bypass turbofan engines to both the community and those on board remains an important issue. At cruise conditions, the jet exit velocity of the bypass flow in many modern turbofans is supersonic. As summarised by \citet{Tam1995}, noise from supersonic jets can be separated into three distinct components: turbulent mixing noise, screech and broadband shock-associated noise (BBSAN). Discrete screech tones are generated by a self-reinforcing feedback loop \citep{powell1953mechanism, raman1999supersonic, edgington2019aeroacoustic}. Non-resonant interaction of jet turbulence with the shock cells produces BBSAN, which is most intense in the sideline directions. Interest in BBSAN remains high for both commercial \citep{huber2014understanding} and high-performance military \citep{vaughn2018broadband} aircraft. This component of supersonic jet noise is the focus of this paper. 

As demonstrated by \citet{harper1973noise}, the broadband noise component is easily identifiable by its directivity and amplitude trends. At higher frequencies, BBSAN is observed to be more dominant than turbulent mixing noise, and its intensity is proportional to the fourth power of the off-design parameter $\beta$, where $\beta^2 = M^2_j-M^2_d$ and the ideally-expanded and design Mach numbers are $M_j$ and $M_d$ respectively. The peak frequency of BBSAN also increases as an observer moves downstream. By modelling the interaction of turbulence with the train of shock cells as a phased array, this frequency trend was successfully reproduced by \citet{harper1973noise}. Their prediction for BBSAN peak frequency $f_p$ is given by
\begin{equation} \label{eq:HBF}
    f_p = \frac{u_c}{L_s(1-M_c\cos \theta)}
\end{equation}
where $u_c$ and $M_c$ are the convection velocity and Mach number of the turbulent structures, $L_s$ is the shock-spacing, and $\theta$ is the angle of observation from the downstream jet axis. The early success of this model substantiated the claim that many features of BBSAN could be explained by the interaction of jet turbulence with the quasi-periodic shock-cell structure.  

BBSAN modelling approaches nonetheless vary. The model developed by \citet{Morris2010} uses solutions of the Reynolds-averaged Navier-Stokes (RANS) equations, requiring only the nozzle geometry and jet operating condition to be specified. Based on an acoustic analogy \citep{Lighthill1952}, construction of the equivalent sources requires turbulent length and time scales which are approximated using the RANS CFD simulations. As the equivalent source behaviour is sensitive to these scales, efforts have been made to refine their description to improve predictions \citep{kalyan2017broad, markesteijn2017supersonic, tan2017application, tan2019correlation}. Within the same framework and by using BBSAN scaling arguments, a different equivalent source term based on decomposing the Navier-Stokes equations was identified by \citet{patel2019statistical}. Reasonable agreement can be obtained with experiments provided the models are calibrated to match the acoustic field.

Rather than focusing on modelling bulk-turbulent statistics, a more fundamental approach was proposed by \citet{Tam1982} on the basis that BBSAN arises from the non-linear interaction between large-scale coherent structures and shocks. The propagating coherent disturbances, resembling the Kelvin-Helmholtz instability in transitional shear layers, motivated the use of linear stability theory \citep{tam1972noise, crighton1976stability}. Hence, the turbulent structures are represented as instability waves \citep{crighton1976stability, tam1979statistical, tam1984sound}, while the periodic shock-cell structure is modelled as a series of time-independent waveguide modes, with wavenumbers $k_n$ and a corresponding shock-cell length approximated by $L_s = 2\pi/k_1$ \citep{Tam1982}. Using this interpretation, $f_p$ can be re-written as 

\begin{equation} \label{eq:HBFtam}
    f_p = \frac{u_ck_n}{2\pi(1-M_c\cos\theta)} ,~n=1,2,3 ...,
\end{equation}
where $n$ is the shock-cell mode. Equation~(\ref{eq:HBFtam}) can also be used to predict peaks generated by higher-order shock-cell modes ($n\geq2$). The work of Tam and co-workers was consolidated into a stochastic model for BBSAN \citep{Tam1987}. Due to the prohibitive cost of the extensive numerical computations required, a similarity source model was constructed which, when compared to experimental measurements \citep{norum1982measurements}, gave favourable noise spectra predictions over a wide range of jet operating conditions. As azimuthally-decomposed BBSAN measurements were not available at the time, scaling coefficients were used to match source model predictions for a single azimuthal mode to the total signal. 

Recently, turbulent mixing noise generation mechanisms in jets have been associated with spatiotemporally coherent structures known as wavepackets. These axially-extended structures have been used extensively for predicting noise radiated from subsonic \citep{Reba2010, cavalieri2012axisymmetric, unnikrishnan2019acoustically}, supersonic \citep{tam1984sound, sinha2014wavepacket} and installed \citep{Piantanida2016} jet flows. A thorough summary on the topic can be found in the review by \citet{Jordan2013}, and the relationship to resolvent modes is discussed in detail by \citet{Cavalieri2019}. The detection of these coherent structures in real flows \citep{suzuki2006instability, kopiev2006experimental, Cavalieri2013,lesshafft2018resolvent}, and our ability to describe them in linearised dynamic models \citep{schmid2002stability, criminale2018theory}, make them ideal candidates to represent the turbulent component of the BBSAN source. The flow properties of large-scale coherent structures, now depicted as wavepackets, may be obtained directly from data \citep{maia2019two}, or alternatively, using solutions to linearised equations with the mean field as a base flow \citep{Cavalieri2013, schmidt2018spectral}. The success of previous studies in using wavepackets to predict far-field noise \citep{lele2005phased} motivates their use to model BBSAN. 

Grounded in stability theory, wavepacket models are well posed and have been used to investigate the underlying sound generation mechanisms for BBSAN. While peak directivity trends were recovered, previous instability wave models for BBSAN offered poor agreement at frequencies above the primary BBSAN peak where sound amplitudes were severely underpredicted \citep{ray2007sound} or artificial dips in the spectra were observed \citep{Tam1987}. The two-point wavepacket model proposed by \citet{Wong2019} offered an explanation. It was shown that, along with higher-order shock-cell modes, coherence decay \citep{cavalieri2014coherence} is essential to broaden the spectral peaks at high frequencies. The inclusion of coherence decay removed the `dips' observed in the predicted acoustic spectra. In \citet{wong2019parabolised}, an equivalent BBSAN source was constructed using PSE to model the wavepackets, along with two-point coherence information derived from an LES database. While a single amplitude scaling coefficient was required to match experimental data, recovery of the spectral shape at high frequencies was encouraging.

In the BBSAN models described above, the `inverse' approach of determining source parameters from the radiated field is ill-posed, as more than one set of parameters may be found to give satisfactory results. Moreover, the parameters found may not be representative of those observed in a real jet. This was indeed explicitly shown by \citet{maia2019two} in modelling source characteristics for a subsonic jet. Using an `inside-out' approach, source parameters, including amplitude, were carefully educed directly from a high-fidelity large-eddy simulation (LES) of a turbulent jet and compared to the parameters previously obtained by \citet{cavalieri2012axisymmetric} for the same inverse problem. Parameter values were clearly shown to differ. An `inside-out' approach was also attempted by \citet{suzuki2016wave} for BBSAN where wavepacket parameters were extracted from the linear hydrodynamic region of an LES database of an underexpanded jet and the shock cells were represented by a number of distinct `Gaussian humps'. The results confirmed modelling assumptions and obtained similar peak predictions to LES results, though agreement at high frequencies remained poor. From these observations, it is evident that a discord remains between the mechanistic insights provided by wavepacket model problems and their ability to accurately predict BBSAN.

This work aims to identify the relevant source mechanisms by extending previous wavepacket-type BBSAN models. This is achieved by using an `inside-out' approach to construct the equivalent source from experimental and numerical flow databases. We adopt the same interpretation of the BBSAN source as \citet{Tam1982} and use Lighthill's acoustic analogy to evaluate the far-field noise. To test the efficacy of the proposed model, sound predictions are compared to the azimuthally-decomposed acoustic data of a target jet case. The source is composed of shock and turbulent components; the shocks are modelled as stationary waveguide modes based on experimental particle image velocimetry (PIV) data. To test which turbulent features are important for sound generation, three descriptions of the wavepackets are obtained, each with an increasing level of complexity. It will be shown that reduced-order linear wavepackets, requiring only a jet mean flow field and a single amplitude parameter, can be used to accurately predict BBSAN peaks across a wide-directivity range. Inclusion of two-point coherence information does indeed recover the `missing sound' at high frequencies. The study we perform is intended to explore the strengths and limitations associated with the use of large-scale coherent structures in BBSAN modelling.

The paper is presented as follows. The mathematical framework for the model is explained in \S~\ref{sec:math1} and the key details of the databases used are outlined in \S~\ref{sec:databases}. We discuss the steps to educe source parameters in \S~\ref{sec:sourceconstruct} and \S~\ref{sec:nearfieldcompare} shows comparisons between simplified flow models with those from the databases for both the shock and turbulent components. We present far-field BBSAN predictions in \S~\ref{sec:results} and source characteristics in \S~\ref{sec:sourcecharacter}. Some conclusions and perspectives are offered in \S~\ref{sec:conclusion}. 

\section{Mathematical formulation} \label{sec:math1}
\subsection{Sound prediction using Lighthill's Acoustic Analogy} \label{sec:math}
The fluctuating sound pressure, $p$, in the acoustic field can be computed using Lighthill's acoustic analogy \citep{Lighthill1952}

\begin{equation}
\label{eq:analogy}
\frac{1}{c^2_\infty }\frac{\partial^2 p}{\partial t^2} - \bigtriangledown^2 p = \frac{\partial^2 T_{ij}}{\partial x_i \partial x_j}, 
\end{equation}
where $t$ is time, $c_\infty$ is the ambient speed of sound, $x$ are the source co-ordinates and $T_{ij}$ is the Lighthill stress tensor

\begin{equation} \label{eq:stress}
    T_{ij} = \rho u_i u_j -\tau_{ij}+(p-c^2_{\infty}\rho)\delta_{ij},
\end{equation}
where $u$ is fluid velocity, $\tau$ are viscous stresses and $\rho$ is density. In high-Reynolds number flows, viscous contribution are minimal \citep{Freund2001} and can hence be neglected. The term $(p-c^2_{\infty}\rho)\delta_{ij}$ represents noise generation due to entropic inhomogeneity. \citet{bodony2008low} have shown that there is significant cancellation between the entropic term and the momentum component ($\rho u_i u_j$) at downstream observer angles in an ideally-expanded supersonic jet. Since BBSAN is dominant in sideline and upstream directions, however, we choose to neglect the entropic term as a first approximation, as it greatly simplifies the model. The stress tensor is hence approximated by

\begin{equation} \label{eq:Tij}
    T_{ij} \approx \rho u_i u_j .
\end{equation}

\begin{figure}
    \centering
    \includegraphics[width=0.6\textwidth]{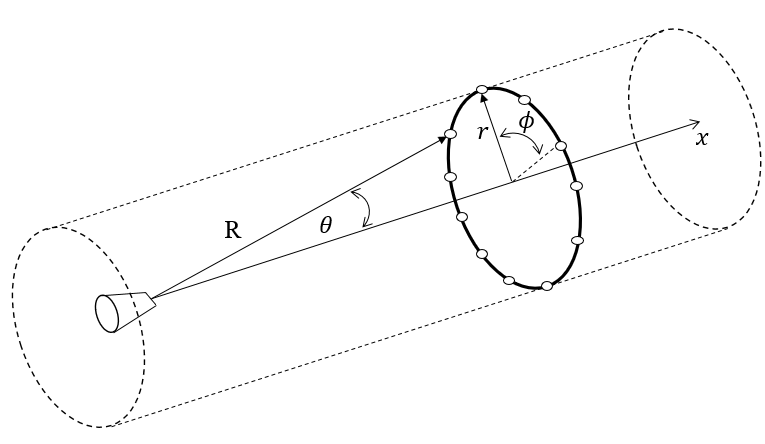}
    \caption{Schematic of experimental setup with the prescribed source $(x,r,\phi)$ and observer $(R,\theta,\phi)$ coordinate systems.}
    \label{fig:coord}
\end{figure}

A solution to equation~(\ref{eq:analogy}) for the acoustic pressure field in the frequency domain, $\omega$, is given by

\begin{equation} \label{eq:psol}
    p(\bm{y};\omega) = \int_V \frac{\partial^2 \hat{T}_{ij}(\bm{x};\omega)}{\partial x_i \partial x_j} G_{0}(\bm{x},\bm{y};\omega) d\bm{x} ,
\end{equation}
where $\hat{T}_{ij}$ is the time Fourier-transformed quantity of $T_{ij}$. An implicit $\exp(-i \omega t)$ dependence on $t$ is assumed. The observer $\bm{y}$ and the source $\bm{x}$ positions are in spherical and cylindrical coordinates respectively as shown in figure~\ref{fig:coord}. The prescribed cylindrical coordinate system $(x,r,\phi)$ has the x-axis aligned with the jet centreline, $r$ is the radial separation and $\phi$ the azimuthal angle. For the observer coordinates $(R,\theta,\phi)$, the same azimuthal coordinate of the cylindrical system is used, the polar angle $\theta$ is defined from the downstream jet axis and $R$ is the distance from the origin. The integration is carried out in the volume $V$ where the source is non-zero. We define $G_0$ as the free-field Green's function 

\begin{equation}\label{eq:green}
G_{0}(\boldsymbol{x},\boldsymbol{y},\omega) = \frac{1}{4 \pi}  \frac{e^{\mathrm{i}k_a | \boldsymbol{x}-\boldsymbol{y} |}}{|\boldsymbol{x}-\boldsymbol{y}|} ,
\end{equation}
where $k_a=\omega/c_\infty$ is the acoustic wavenumber. We also transfer the second derivative of $T_{ij}$ onto the Green's function by applying the divergence theorem and assuming the resulting surface integral to be negligible \citep{goldstein1976aeroacoustics}. This makes evaluation of the integral less sensitive to spurious fluctuations in the stress tensor due to numerical noise.

Since flow fluctuations are not square-integrable functions, as required for the application of a Fourier transform, one cannot obtain the sound field through direct application of equation~(\ref{eq:psol}). One way to circumvent this issue \citep{landahl1989turbulence,cavalieri2014coherence, baqui2015coherence} is to compute the power spectral density (PSD) of the acoustic field. For a given frequency $\omega$, the PSD $\langle p(\bm{y},\omega)p^*(\bm{y},\omega) \rangle$ is given by

\begin{eqnarray} \label{eq:psd1}
    &&\langle p(\bm{y};\omega)p^*(\bm{y};\omega) \rangle  =   \nonumber\\
    && \int_V \int_V \langle T_{ij}(\bm{x_1};\omega)T^*_{ij}(\bm{x_2};\omega)\rangle \frac{\partial^2 G_{0}(\bm{x_1},\bm{y};\omega)}{\partial x_i \partial x_j} \frac{\partial^2 G^*_{0}(\bm{x_2},\bm{y};\omega)}{\partial x_i \partial x_j} d\bm{x_1}d\bm{x_2} ,
\end{eqnarray}
where $\langle \rangle$ denotes an expected value, the quantity $\langle T_{ij}(\bm{x_1},\omega)T^*_{ij}(\bm{x_2},\omega)\rangle$ is the cross-spectral density (CSD) of the stress tensor for a pair of points $\bm{x_1}$ and $\bm{x_2}$, $^*$ denotes the complex conjugate and we have dropped the hats for convenience. We exploit axisymmetry by expanding $T_{ij}$ as a series of azimuthal modes \citep{ michalke1975turbulence}; noting that there is a direct correspondence between the azimuthal mode of the source and that of the sound field \citep{michalke1970wave, cavalieri2012axisymmetric}. By taking a Fourier transform of the source in azimuth, we can compute azimuthal mode $m$ of the far-field pressure to be

\begin{eqnarray} \label{eq:psd2}
    &&\langle p(R,\theta;m,\omega)p^*(R,\theta;m,\omega) \rangle  =   \nonumber\\
    && \int_V \int_V \langle S_{ij}(m,\omega)\rangle \frac{\partial^2 G_{0,1}(m,\omega)}{\partial x_i \partial x_j} \frac{\partial^2 G^*_{0,2}(m,\omega)}{\partial x_i \partial x_j} d\bm{x_1}d\bm{x_2} , 
\end{eqnarray}
where we have dropped the spatial coordinates of the source for compactness, $G_{0,1}$ and $G_{0,2}$ represent the Green's functions at source location $\bm{x_1}$ and $\bm{x_2}$ respectively, and $S_{ij}$ represents the CSD of the stress tensor
\begin{equation} \label{eq:sij}
    S_{ij}(x_1,r_1,x_2,r_2;m,\omega) = T_{ij}(x_1,r_1;m,\omega)T^*_{ij}(x_2,r_2;m,\omega) .
\end{equation}

\subsection{Equivalent BBSAN source model}
The proposed BBSAN model is based on the idea that the source only involves fluctuations associated with interactions between the turbulent component ($\bm{q}_t$) and shock perturbations ($\bm{q}_s$). This assumption has been made by a number of authors \citep{Tam1982,lele2005phased,ray2007sound,Wong2019}, where different descriptions of $\bm{q}_t$ and $\bm{q}_s$ were investigated. We follow this approach and, similar to \citet{Wong2019}, adopt a two-point description of the source. 

As performed by \citet{Tam1987}, we decompose the flow variables according to

\begin{equation}\label{eq:decomp}
    \bm{q} = \bm{\bar{q}} + \bm{q}_t + \bm{q}_s ,
\end{equation}
where $\bm{\bar{q}}$, $\bm{q}_t$, $\bm{q}_s$ are the mean, turbulent and shock-cell disturbance components respectively. We take the mean component to be the time-averaged flow of an ideally-expanded jet. The vector $\bm{q}$ refers to the dependent flow variables of interest, $\bm{q} = [u_x, u_r, u_\phi, T, \rho]^T $, where $u_x$, $u_r$ and $u_\phi$ are the axial, radial and azimuthal velocity components respectively. The thermodynamic variables include $T$ and $\rho$ which are the temperature and the density of the fluid respectively. The decomposition in equation~(\ref{eq:decomp}) is substituted into the stress tensor in equation~(\ref{eq:Tij})

\begin{equation} \label{eq:subTij}
    T_{ij} \approx (\bar{\rho}+\rho_s+\rho_t)( \bar{u}_i+u_{i,t}+u_{i,s})( \bar{u}_j+u_{j,t}+u_{j,s}) .
\end{equation}
Assuming that BBSAN is generated by turbulence-shock interaction, the expression for $T_{ij}$, as shown in appendix~\ref{app:A}, can be simplified to

\begin{equation} \label{eq:BBSANTij}
T_{ij} \approx \bar{\rho}(u_{i,t}u_{j,s}+u_{i,s}u_{j,t})+\rho_s(\bar{u}_iu_{j,t}+\bar{u}_ju_{i,t})+\rho_t(\bar{u}_iu_{j,s}+\bar{u}_ju_{i,s}) .
\end{equation}
Here we highlight some characteristics of equation~(\ref{eq:BBSANTij}). Firstly, this representation of $T_{ij}$ does not account for turbulent mixing noise since only turbulence-shock interaction terms are retained (appendix~\ref{app:A}). This is justified by the minimal contribution of mixing noise at the frequencies and polar positions where BBSAN is dominant \citep{Viswanathan2006, viswanathan2010characteristics}. Agreement with measured acoustic data at low frequencies and downstream polar angles would therefore not be expected. Secondly, unlike previous wavepacket models in subsonic jets \citep{cavalieri2011jittering, Piantanida2016, maia2019two}, we retain all velocity components of $T_{ij}$ in order to improve predictions in the sideline direction. We also note that while equation~(\ref{eq:BBSANTij}) is similar to the source term derived by \citet{lele2005phased}, we retain the double-divergence and have discarded the entropic term. 

The BBSAN sound field can be obtained using equations~(\ref{eq:psd2}) and~(\ref{eq:BBSANTij}). Unlike previous two-point wavepacket modelling work \citep{maia2019two, Wong2019}, we choose to drop the line-source simplification and work with a full volumetric source instead. The $\bm{q}_t$ and $\bm{q}_s$ parts of $T_{ij}$ are each computed using numerical and experimental databases, respectively, as shown in \S~\ref{sec:sourceconstruct}, before being combined according to equation~(\ref{eq:BBSANTij}). A spatial Hann window is used to smoothly truncate the source domain ($0\leq x \leq 15D$) to ensure no artificial overprediction of the acoustic field \citep{martinez2008correction}. The radial domain extends to $r=2D$ while the downstream bound of the domain is chosen based on the negligible contributions of the equivalent source past $x=10D$ (\S~\ref{sec:shockcompare}).

\section{Databases} \label{sec:databases}
To explore the sound source mechanisms, far-field acoustic spectra predictions are computed and compared to experimental measurements. The goal is to build an equivalent source appropriate for describing the sound field for a target jet operating condition. The model is based on a decomposition of the flow field into $\bm{\bar{q}}$, $\bm{q}_t$ and $\bm{q}_s$ components (equation~(\ref{eq:decomp})).

We obtain this data from different databases; wavepackets are educed from an ideally-expanded jet, while the modelling of the shock disturbances is based on an underexpanded jet. Ideally, the exit conditions of these jets (NPR, $M_j$, $Re$, $T_j$) should be as close as possible to the target case. 

The flow-field databases are summarised in \S~\ref{sec:LESdata}-\ref{sec:piv} while the acoustic measurements of the target jet are described in \S~\ref{sec:acousticdata}. A summary of the jet operating conditions is provided in table~\ref{tab3}. We note that the databases do not correspond to identical operation conditions. They are here only used to inform our modelling choices such that the descriptions of $\bm{q}_t$ and $\bm{q}_s$ align closely with a realistic jet. Given the small discrepancies between the databases, we perform a short sensitivity study to assess how these may impact BBSAN peak frequency and amplitude. This is provided in appendix~\ref{sec:note}. 

\begin{table}
\centering 
\begin{tabular}{c c c c c c c c c } 
Database &$M_j$ & $M_d$ &  $NPR$ &   $T_j/T_{\infty} $ &   $D_j/D$ &   $Re$ & $\beta$  \\ 
LES & 1.50  &1.5 &3.67 &1.0 &1.0 &\centering $1.76 \times 10^6$& 0 \\
PIV &1.45&1.0  &3.40 &0.70 &1.07 &$8.51 \times 10^5$ &1.05 \\ 
Acoustic & 1.50 &1.0  &3.67 &0.69 &1.09 &$4.50\times10^5$ &1.12 \\ 
\end{tabular}
\caption{Summary of jet operating parameters for each database.} \label{tab3}
\end{table}

\subsection{Numerical database: Large-eddy simulation of $M_j=1.5$ ideally-expanded jet} \label{sec:LESdata}
The turbulent flow quantities $\bm{q}_t$ are extracted from a large-eddy simulation (LES) of an isothermal ideally-expanded $M_j=1.5$ supersonic jet. An extension to the previous LES by \citet{bres2017unstructured}, this simulation was performed using the
compressible flow solver ``Charles", developed at Cascade Technologies, on an unstructured adapted grid with 40 million cells. The jet issues from a round converging-diverging nozzle. The Reynolds number based on nozzle exit conditions is $Re = \rho_j U_j D / \mu_j = 1.76 \times 10^6$, matching the experiment carried out at the United Technologies Research Center (UTRC) anechoic jet facility \citep{schlinker2009supersonic}. Near-wall adaptive mesh refinement is employed on the internal nozzle surface to closely model the boundary layer inside the nozzle, leading to turbulent boundary layer profiles at the exit \citep{bres2018importance}. A slow co-flow of $M_{co} = 0.1$ is also included in the simulation to match the UTRC experimental conditions. 

To facilitate post-processing and analysis, the LES data is interpolated from the original unstructured LES grid onto a structured cylindrical grid with uniform spacing in azimuth. The three-dimensional cylindrical grid is defined over $0 \leq x/D \leq  30 $, $0 \leq   r/D \leq 6$, with $(n_x, n_r, n_{\theta} ) = (698,  136,  128)$, where $n_x$, $n_r$ and $n_{\theta}$  are the number of grid points in the streamwise, radial and azimuthal direction, respectively. The simulation time step, in acoustic time units, is  $\Delta t c_\infty /D  = 0.0004$ and the database is sampled every $\Delta t c_\infty /D = 0.1$. Snapshots are therefore recorded every 250 time steps, corresponding to a cutoff (Nyquist) frequency of $St = \Delta f D / U_j =  3.33$. The simulation parameters are summarised in table~\ref{tab1}. Further details on the numerical strategy can be found in \citet{bres2017unstructured}.

\begin{table}
\centering 
\begin{tabular}{c c c c c} 
 [$n_x,n_r,n_\theta$]&Sim. Duration & Sampling Period &Nyquist Freq. & Num. Snapshots\\ 
$698, 136, 128$& 1000 &0.1 & 3.33 & 10000 \\ 
\end{tabular}
\caption{Summary of LES parameters.} \label{tab1}
\end{table}

\subsection{Experimental database: Particle image velocimetry of $M_j=1.45$ underexpanded jet} \label{sec:piv}
For the description of $\bm{q}_s$, we resort to high spatial resolution 2D 2C particle image velocimetry (PIV) measurements of a cold screeching underexpanded supersonic jet with an ideally-expanded Mach number of $M_j=1.45$. The data was previously acquired at the supersonic jet facility at the Laboratory for Turbulence Research in Aerospace and Combustion (LTRAC) \citep{edgington2014underexpanded}. The facility has been used extensively in previous experimental studies of shock-containing supersonic jets \citep{edgington2014coherent, weightman2019nozzle}. The facility is not anechoic and noise measurements were not conducted. 

\begin{figure}
    \centering
    \includegraphics[width=1.0\textwidth]{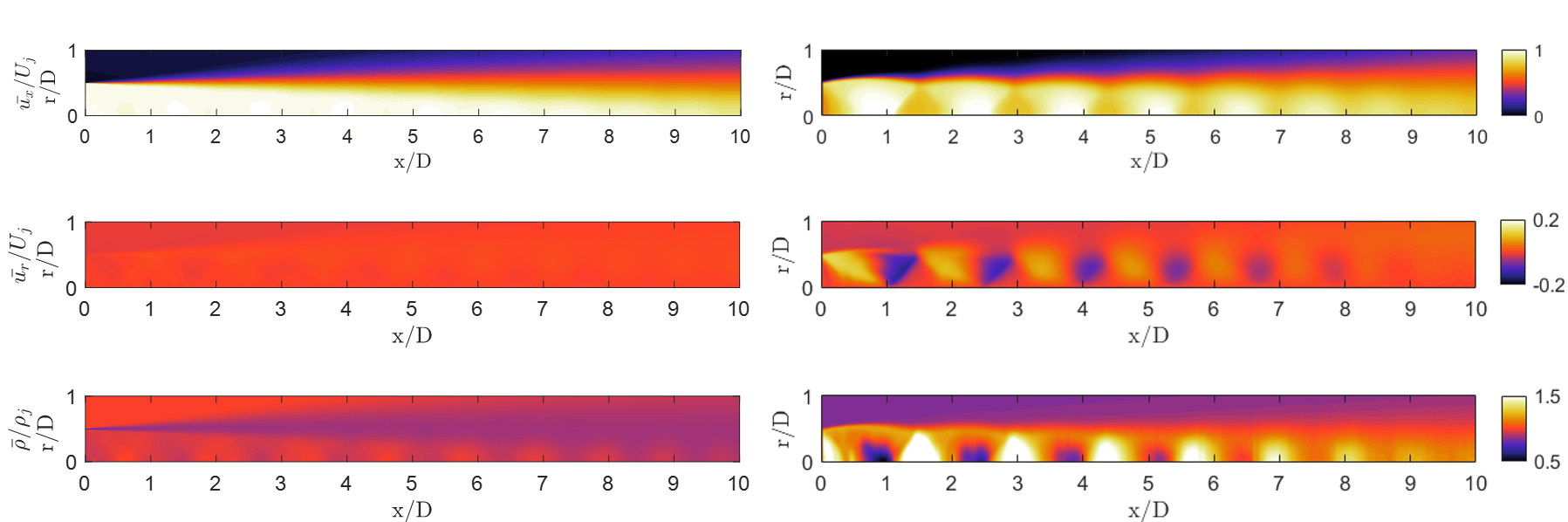}
\caption{LES (left) and PIV (right) $x-r$ contour mean fields for ideally-expanded and shock-containing jets respectively; streamwise velocity (top), radial velocity (centre) and density (bottom). Flow quantities are normalised by the ideally-expanded condition. }\label{fig:meanfield}
\end{figure}

The final field of view of the images is $10D$ and $2.2D$, with $(N_x,N_y)= (1000,75)$, in the axial and radial directions respectively. The optical resolution of the images is $0.001D/px$. Full details of the experimental set-up and post-processing techniques are described in \citet{edgington2014coherent}. Mean axial and radial velocity fields from both the LES and PIV data are shown in figure~\ref{fig:meanfield}.

\subsection{Acoustic database: Far-field acoustic measurements $M_j=1.5$ underexpanded jet } \label{sec:acousticdata}
The acoustic measurements were performed at the Supersonic Jet Anechoic Facility (SJAF) at Monash University. This is a different facility to the jet rig used to acquire the PIV measurements in \S~\ref{sec:piv}. Most importantly, the jet is mounted inside a fully-enclosed anechoic chamber. The chamber walls are treated with 400mm foam wedges, corresponding to a cut-off frequency of 500Hz. The interior chamber dimensions (wedge-tip-to-wedge-tip) are 1.5m $\times$ 1.2m $\times$ 1.4m. The jet exits out of a converging-round nozzle with an exit diameter of $D=8$mm. Unheated compressed air is supplied to the jet at $NPR=3.67$, corresponding to the same $M_j$ as the LES case. 

Acoustic measurements were performed using an azimuthal ring of radius $11D$ and a schematic of the experimental setup is shown in figure~\ref{fig:coord}. The CSD of pressure, as a function of azimuthal separation, was obtained using a pair of G.R.A.S. Type 46BE 1/4" pre-amplified microphones with a frequency range of 4Hz-100kHz, one fixed and the other moving in the azimuthal direction. Using the post-processing methodology detailed in \citet{wong2020azimuthal}, the measured sound fields were azimuthally decomposed. The azimuthal array is traversed axially to acquire measurements at different polar angles over a cylindrical surface. The radial distance $r=11D$ is therefore constant, while observer distance $R$ changes. A detailed description of the experimental setup can be found in \citet{wong2020azimuthal}.

The motivation for using azimuthally-decomposed data is twofold. Firstly, the measurements of previous authors \citep{suzuki2016wave, arroyo2019azimuthal, wong2020azimuthal} suggest the spectrum of each azimuthal mode differs from the total sound field; an increasing number of modes is required to reconstruct the total signal at high frequencies and for upstream angles. Secondly, in a linear acoustic problem such as this, \citet{michalke1975turbulence} demonstrated there exists a direct correspondence between the acoustic source $S_{ij}$ and the far-field sound of the same azimuthal mode.

\section{Construction of source variables} \label{sec:sourceconstruct}
This section details the procedures used to compute the source variables in equation~(\ref{eq:decomp}) using the databases described in the preceding section. Each source variable ($\bm{\bar{q}}$, $\bm{q}_t$ and $\bm{q}_s$) is either obtained via direct substitution of LES data or constructed using models informed by flow information from the LES and PIV databases.

\subsection{Eduction of shock-cell component} \label{sec:educeshock}
Similar to \citet{Tam1982} and \citet{lele2005phased}, we adopt the Pack and Prandtl \citep{prandtl1904stationaren, pack1950note} approximation of the shock-cell structure. The shocks are modelled as small disturbances superimposed over an ideally-expanded jet. The model assumes the jet to be bounded by a vortex sheet, allowing the periodic shock-cell structure to be represented by a sum of zero-frequency waves. Good agreement is found close to the nozzle exit, where the shear layer is thin, but worsens downstream as the shear layer thickens, invalidating the vortex sheet assumption \citep{tam1985multiple}. With increasing distance from the nozzle exit, the model therefore fails to predict the decay in shock strength and the accompanying contraction in shock-cell spacing. Since the BBSAN source is reported to extend several several jet diameters downstream \citep{seiner1980aerodynamic,gojon2017numerical}, any disagreement between the vortex sheet model and measured jet characteristics is likely to lead to incorrect peak frequency predictions. 

While the shock-cell disturbances may be extracted from data (e.g. PIV) or computed by solving linear locally-parallel stability equations \citep{tam1985multiple}, the shock perturbations have a smooth and nearly sinusoidal variation towards the end of the potential core. The Pack and Prandtl (P-P) model therefore remains an attractive simplified approach for capturing the mean shock structure; indeed, source models adopting the approximation are able to reproduce the main features of BBSAN, including higher-order BBSAN peaks \citep{Tam1982, Wong2019}. To remedy the shortfalls of the vortex sheet assumption, we use the PIV database to modify the P-P solution in order to arrive at a more realistic model.   

The jet is modelled as a cylindrical vortex sheet \citep{lessen1965instability}, and the normal mode Ansatz is introduced

\begin{equation}
   \bm{q}_{s,vortex}(x,r,\theta,t) = \sum_{\omega} \sum_{k_s} \sum_{m_s} \bm{\hat{q}}_s(r)e^{i\omega_st - ik_sx - im_s\phi} ,
\end{equation}
where $\omega_s$ is frequency, $k_{s}$ and $m_s$ are axial and azimuthal wavenumbers. By assuming the shock-cell disturbances are stationary ($\omega_s=0$) and axisymmetric ($m_s=0$), we obtain for each dependent variable of interest $\bm{q}_s$,

\begin{equation}
    q_{s,vortex}(x,r) = \sum^\infty_{n=1}A_n J_0(\alpha_nr)e^{-ik_{s_n}x} ,
\end{equation}
where $A_n$ is the amplitude of each shock-cell mode $n$, $k_{s_n}$ are the axial wavenumbers and $J_0$ is the zeroth-order Bessel function of the first kind. The boundary condition for constant velocity on the jet boundary \citep{pack1950note} requires that the values of $\alpha_n$ satisfy

\begin{equation}
    J_0(\alpha_n) = 0 ,
\end{equation}
and from the dispersion relation, we obtain the sequence of axial wavenumbers to be
\begin{equation}
    k_{s_n} = \frac{\alpha_n}{\sqrt{M^2_j-1}} .
\end{equation}

In real jets, $A_n$ and $k_{s_n}$ are functions of $x$, as the underlying evolution of the mean flow modifies each Fourier component. This variation is not captured in the P-P model due to the parallel vortex-sheet assumption. Hence, we wish to obtain a modified version of the vortex sheet model, $\bm{q}_{s,mod}$, which more closely resembles measured shock-containing jet characteristics. A realistic representation of $\bm{q}_s$ is obtained by subtracting the ideally-expanded flow quantities of the LES dataset from the shock-containing quantities of the PIV dataset 

\begin{equation}\label{eq:deccomp1}
    \bm{q}_s \approx \bm{q}_{PIV} - \bm{q}_{LES} ,
\end{equation}
where we have assumed the quantity $\bm{q}_{LES}$ contains both the mean and turbulent contribution in (\ref{eq:decomp}). While the PIV data provides axial and radial velocities, the mean shock-associated density modulation ($\rho_s$) is estimated using the ideal gas law, with reconstructed temperatures and pressures obtained by the method of \citet{tan2018equivalent}. Good agreement is observed between the reconstructed densities and mean background-oriented schlieren (BOS) measurements \citep{tan2015measurement}. LES quantities are then interpolated onto the lower-resolution PIV grid. 

To adjust $k_{s_n}$, a Fourier transform of $\bm{q}_s$ is performed downstream of the nozzle exit to capture the variation of shock-cell spacing, similar to \citet{Morris2010}. The axial wavenumber from the vortex-sheet approximation is adjusted empirically, using a linear fit to match the PIV data

\begin{equation} \label{eq;shockspacing}
    k_{s_n, mod} = 0.79\times k_{s_n,vortex}+1.02 .
\end{equation}

To determine the axial variation in $A_n$, we assume there exists a relationship between the vortex sheet model $\bm{q}_{s,vortex}$ and the adjusted values $\bm{q}_{s,mod}$ 

\begin{equation}
    \bm{q}_{s, mod}(x,r;n) = b(x;n)\bm{q}_{s,vortex}(x,r;n) ,
\end{equation}
where the factor $b(x;n)$ is determined by using the experimentally-deduced values $\bm{q}_s$,

\begin{equation}
    b(x;n) = \frac{\langle \bm{q}_{s,vortex}(x,r;n),\bm{q}_{s}(x,r) \rangle }{\left|| \bm{q}_{s,vortex}(x,r;n) \right||^2},
\end{equation}
and the inner-product is defined as

\begin{equation}
    \langle \bm{q}_{s,vortex}(x,r;n),\bm{q}_{s}(x,r) \rangle = \int^{R}_0 \bm{q}_{s,vortex}(x,r';n)\bm{q}^*_{s}(x,r')\bm{W}(x,r') r' dr', 
\end{equation}
where the orthogonality of Bessel functions is exploited. We also assign null weights to the temperature component of $\bm{W}$, since we are only concerned with the density and velocity components that contribute to the BBSAN source term in (\ref{eq:BBSANTij}). The integration limit $R$ is taken to be the maximum radius of the PIV measurement domain. 

Unlike \citet{ray2007sound}, higher-order modes ($n>1$) are included in our shock-cell description. \citet{Wong2019} used a line-source wavepacket model, incorporating the effects of coherence decay, to demonstrate the importance of higher-order modes at high frequencies, despite the fact they possess wavenumbers which lie outside the radiating range \citep{ray2007sound}. The final shock-cell structure is reconstructed using three modes ($n=1,2,3$), as this was deemed suitable for predicting the far-field BBSAN over the frequency range of interest.

\subsection{Eduction of wavepacket component}
Two methods are used to obtain the turbulent (wavepacket) component of the source $T_{ij}$. The first method involves the direct substitution of post-processed LES data, representing the most `complete' prediction possible for the proposed BBSAN model as it encapsulates the full range of resolved spatial and temporal turbulent scales. The second utilises solutions to parabolised stability equations (PSE), which have previously been shown to be appropriate reduced-order representations of the large-scale perturbations in turbulent jets \citep{gudmundsson2011instability, Cavalieri2013, sinha2014wavepacket}.  

\subsubsection{LES database}
The LES data contains a broad range of temporal and spatial scales. To handle this, extraction of coherent wavepackets is performed in a similar fashion to previous studies \citep{sinha2014wavepacket,schmidt2017wavepackets,maia2019two}, assuming the jet to be periodic in azimuth ($\phi$) and statistically stationary. The fluctuating turbulence variables $\bm{q}_t$ are decomposed using the following ansatz,

\begin{equation} \label{eq:ansatz}
    \bm{q}_t(x,r,\phi,t) = \sum_\omega \sum_m \hat{\bm{q}}_t(x,r)e^{-i\omega t+im\phi} ,
\end{equation}
where $\omega$ is angular frequency and $m$ is azimuthal wavenumber of the wavepacket. Using this decomposition, the LES data is Fourier-transformed in both azimuth and time. For each azimuthal mode $m\neq 0$, the contribution from the positive mode $+m$ is combined with the complex conjugate of that from the negative mode $-m$, since the jet has no swirl. Prior to the temporal Fourier transform, the time series is divided into data blocks of $N_{fft}=128$ sample points and a Hann window is applied to suppress spectral leakage. The final number of blocks is $N_B=310$, with a 75\% overlap, was sufficient to ensure statistical convergence. For a given $\omega$ and $m$, the $\mathcal{J}^{th}$ block of the Fourier-transformed flow field $\bm{q}^{(\mathcal{J}^{th})}_{m,\omega}$ is obtained and substituted directly into the $\bm{q}_t$ part of $T_{ij}$ in (\ref{eq:sij}). Fluctuations extracted from the LES data do not undergo any additional processing. The $\bm{q}_{t}$ (from LES) and $\bm{q}_{s}$ (from PIV) parts are then combined to produce the BBSAN source term, given by

\begin{equation} \label{eq:sourcecomples}
   S_{ij, LES}(\bm{x_1},\bm{x_2};m,\omega) = \frac{1}{N_B} \sum^{\mathcal{J}=N_B}_{\mathcal{J}=1}  T^{(\mathcal{J})}_{ij}(\bm{x_1};m,\omega)T^{*(\mathcal{J})}_{ij}(\bm{x_2}; m,\omega) .
\end{equation}

\subsubsection{Parabolised stability equations}
The use of PSE to model wavepackets has been well studied in both subsonic \citep{gudmundsson2011instability, Cavalieri2013} and supersonic \citep{sinha2014wavepacket, rodriguez2015study, kleine2017evaluation} turbulent jets where the mean flow is assumed to be slowly diverging. The PSE approach has also been used to model the turbulent component in previous BBSAN models \citep{ray2007sound, wong2019parabolised}. 

The PSE system follows the same non-dimensionalisation and ansatz (\ref{eq:ansatz}) used to decompose the LES data. It is assumed that $\bm{q}_t(x,r,\phi,t)$ may further be decomposed into a slowly and rapidly varying component. The appropriate multiple-scales ansatz, first proposed by \citet{crighton1976stability}, can be written

\begin{equation} \label{eq:pse}
\bm{q}_t(x,r,\phi,t) = \hat{\bm{q}}_t(x,r)e^{\mathrm{i}\int \alpha(x')dx'}e^{-\mathrm{i}\omega t}e^{\mathrm{i}m\phi},
\end{equation}
where the rapidly and slowly-varying parts are described by the exponential term $e^{\mathrm{i}\int \alpha(x')dx'}$, and the modal shape function $\hat{\bm{q}}_t$, respectively. The integrand $\alpha(x')$ is the complex-valued hydrodynamic wavenumber that varies with axial position. Equation~(\ref{eq:pse}) can be substituted into the governing inviscid linearised equations. The resultant matrix system is recast into the following compact form

\begin{equation}
\bm{A}\hat{\bm{q}}_t+\bm{C}\frac{\partial \hat{\bm{q}}_t}{\partial x} + \bm{D}\frac{\partial \hat{\bm{q}}_t}{\partial r} = 0 ,
\end{equation}
where the left-hand side is the linear operator acting on a given $(m,\omega)$ shape function $\hat{\bm{q}}_t$. Full expressions for operators $\bm{A}, \bm{C}$ and $\bm{D}$ can be found in \citet{fava2019propagation}. To find $\alpha(x)$ and $\hat{\bm{q}}_t$, the system is discretised and solved by streamwise spatial marching. Chebyshev polynomials are used to  discretise the radial domain and first-order finite differences to approximate the axial derivatives. The axial step-size $\Delta x$ is limited by the numerical stability condition specified by \citet{li1997spectral}

\begin{equation} \label{eq:psestep}
    \Delta x \geq \frac{1}{|Re\left \{\alpha_{m,\omega}(x)\right \}|}.
\end{equation}

As discussed by \citet{herbert1997parabolized} and \citet{Cavalieri2013}, there remains an ambiguity in the PSE decomposition, since the spatial growth of $\bm{q}_t$ is shared by both the shape function $\hat{\bm{q}}_t$ and the complex amplitude $e^{\mathrm{i}\int \alpha(x')dx'}$. A normalisation condition is introduced to remove this ambiguity

\begin{equation}
    \int^\infty_0 \hat{\bm{q^*}_t}\frac{\partial \hat{\bm{q}}_t}{\partial x} rdr = 0 .
\end{equation}

Dirichlet boundary conditions are used as $r\rightarrow\infty$ and the condition along the jet centreline follows the treatment prescribed in \citet{mohseni2000numerical} using parity functions. A complete description of the procedure is provided by \citet{gudmundsson2011instability} and a good summary can be found in \citet{sasaki2017real}.

The PSE solutions are computed using the the mean flow of the ideally-expanded jet LES. The LES mean flow is linearly interpolated onto the PSE grid, and for each frequency, solved on its own axial grid given by the minimum step-size specified in equation~(\ref{eq:psestep}). To initiate the marching procedure, initial flow conditions at the nozzle exit plane are provided by the eigenfunction of the Kelvin-Helmholtz (KH) instability mode, obtained by solving the locally-parallel stability problem. 

Wavepacket amplitudes are undefined, as PSE solves a linear problem. For meaningful comparisons, PSE solutions must be scaled to experimental results. Different approaches to the task have been performed by previous authors and a summary is provided by \citet{rodriguez2015study}. Method complexity ranges from a simple scalar multiplication, to more robust bi-orthogonal projections of LES data onto PSE wavepackets near the nozzle exit \citep{rodriguez2013inlet}. While PSE scaling approximately follows an exponential trend with frequency \citep{antonialli2018amplitude}, scaling amplitudes are found to be sensitive to the choice of the matching flow variables, region of interest and the axial position.

A scaling method compatible with the goal of this study, that is, to develop a BBSAN model that does not require calibration from far-field acoustic data, demands that the amplitude of the source term must be obtained directly from the flow information. This requires the PSE solution to be scaled to the same amplitude as the extracted LES fluctuations. The most stringent method obtains the PSE amplitudes based solely on flow-field quantities of the LES data at a single given axial station $x_0$. We define the source-based inner product of the PSE solutions $\bm{q}_{t,PSE}$ and the $\mathcal{J}^{th}$ block of the processed LES data $\bm{q}^{(\mathcal{J}^{th})}_{t,LES} $ as

\begin{eqnarray} \label{eq:PSEinnerproduct}
    &&\langle \bm{q}_{t,PSE}(x,r;m,\omega),\bm{q}^{(\mathcal{J}^{th})}_{t,LES}(x,r;m,\omega) \rangle = \\
    &&\int^{R}_0 \bm{q}_{t,PSE}(x,r;m,\omega)\bm{q}^{*(\mathcal{J}^{th})}_{t,LES}(x,r;m,\omega) \bm{W}(x,r') r' dr', \nonumber
\end{eqnarray}
where we have again assigned null weights to the temperature component, and $R$ is determined by the outer bound of the LES data. We assume the LES flow variables may be expressed in the form
\begin{equation}
    \bm{q}^{(\mathcal{J}^{th})}_{t,LES}(x,r;m,\omega) = \mathcal{A}(x;m,\omega)\bm{q}_{t,PSE}(x,r;m,\omega) ,
\end{equation}
where the value $\mathcal{A}$ is evaluated for every $\mathcal{J}^{th}$ block according to
\begin{equation} \label{eq:betascale}
    \mathcal{A}^{(\mathcal{J}^{th})}(x;m,\omega) = \frac{\langle \bm{q}_{t,PSE}(x,r;m,\omega),\bm{q}^{(\mathcal{J}^{th})}_{t,LES}(x,r;m,\omega)  \rangle }{\left|| \bm{q}_{t,PSE}(x,r;m,\omega) \right||^2} .
\end{equation}
For each frequency-azimuth pair, the axial scaling location is chosen to be the peak of the PSE wavepacket $x_0$; $\mathcal{A}(x_0;m,\omega)$ becomes the PSE scaling factor. Each value of $\mathcal{A}$ is averaged over the total number of blocks $N_B$. The scaled PSE solutions are then substituted into the turbulent part of equation~(\ref{eq:BBSANTij}) and a statistical, perfectly-coherent BBSAN source, $\check{S}_{ij}$, is given by

\begin{equation} \label{eq:sourcecomples1}
   \check{S}_{ij,PSE}(\bm{x_1},\bm{x_2};m,\omega) =  \check{T}_{ij}(\bm{x_1};m,\omega)\check{T}^{*}_{ij}(\bm{x_2}; m,\omega).
\end{equation}

\subsection{Coherence-matched source term}
For a BBSAN line-source model, \citet{Wong2019} demonstrated that the use of wavepacket solutions from PSE gives rise to non-physical dips in the far-field sound spectrum. This is due to the PSE-derived wavepackets, and hence the statistical source $\check{S}_{ij}$, having unit coherence between any pair of points \citep{cavalieri2014coherence}. Instead, two-point coherence information of the flow field, which represents randomness in wavepacket phase statistically \citep{cavalieri2011jittering}, smooths out higher-order BBSAN peaks and results in the recovery of missing sound at upstream angles. To reproduce the original source $S_{ij}$, in addition to amplitude and phase velocity, two-point coherence of the source must also be matched \citep{cavalieri2014coherence, maia2019two}. The CSD of $S_{ij}$ becomes

\begin{equation}
    \langle T_{ij}(\bm{x_1};m,\omega)T^*_{ij}(\bm{x_2};m,\omega)\rangle = \gamma^2(\bm{x_1},\bm{x_2};m, \omega)  \check{T}_{ij}(\bm{x_1};m,\omega)\check{T}^*_{ij}(\bm{x_2};m,\omega),
\end{equation}
where $\gamma$ is the coherence between two points $\bm{x_1}$ and $\bm{x_2}$. Unlike previous studies \citep{baqui2015coherence, maia2019two}, we do not model the coherence envelope but, rather compute it directly from the LES data. The coherence profile of $S_{ij,LES}$ (equation~(\ref{eq:sourcecomples})) is computed between all sets of points in the source region, given by

\begin{equation} \label{eq:Gamma}
\gamma^2 (\bm{x_1},\bm{x_2};m, \omega) =\frac{|\langle S_{ij,LES}(\bm{x_1},\bm{x_2};m, \omega)\rangle|^2}{\langle|S_{ij,LES}(\bm{x_1})|^2\rangle \langle|S_{ij,LES}(\bm{x_2})|^2\rangle} .
\end{equation}

\subsection{Summary of BBSAN source model construction} \label{sec:summaryconstruct}
\begin{figure}
    \centering
    \includegraphics[width=1.0\linewidth]{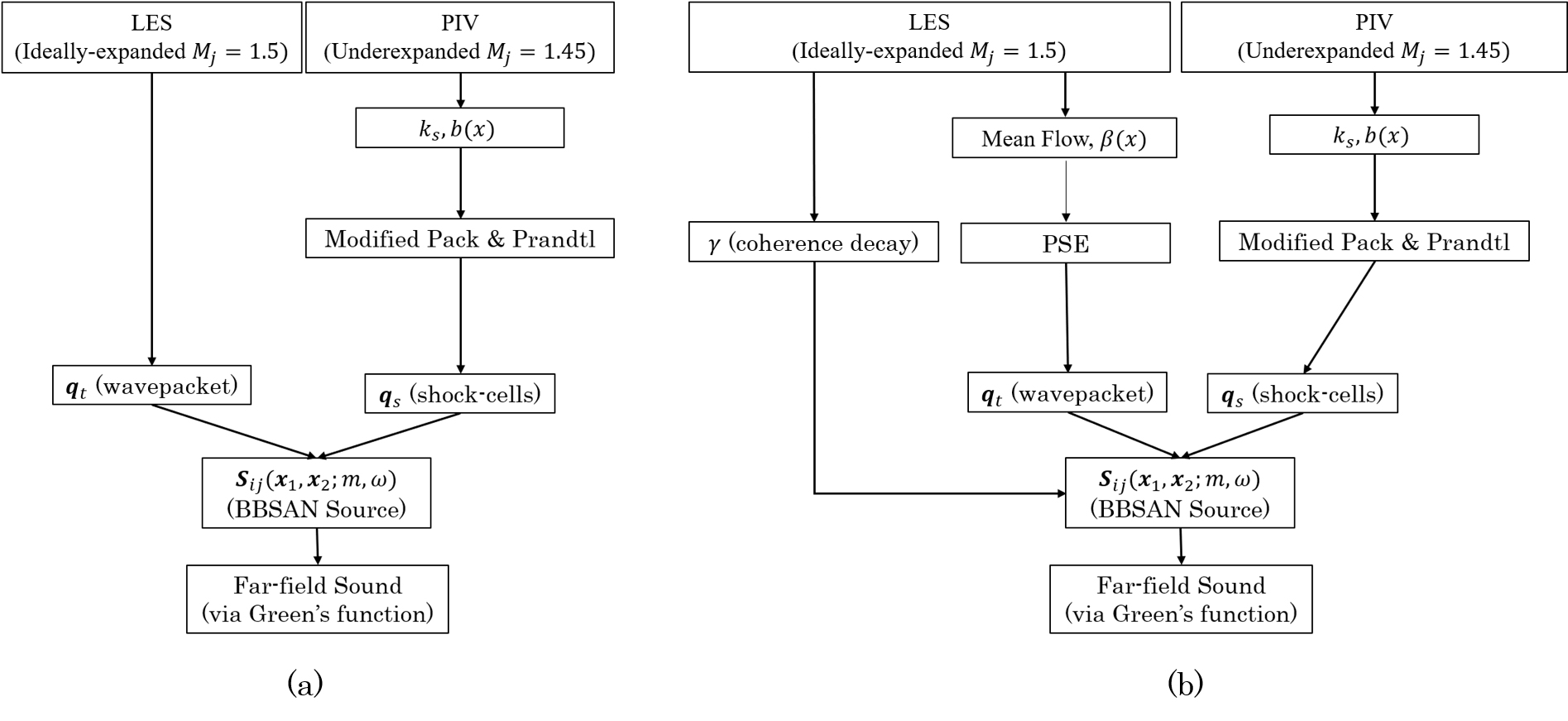}
    \caption{Summary of BBSAN model construction; (a) source model with $\bm{q}_t$ obtained directly from LES data and (b) statistical source model with $\bm{q}_t$ obtained from PSE solutions.}
    \label{fig:modelconstruct}
\end{figure}

An overview of the BBSAN source assembly is shown in figure~\ref{fig:modelconstruct}. From here, we shall refer to the `LES model' where wavepacket fluctuations are extracted directly from LES data (figure~\ref{fig:modelconstruct}a) and the `PSE model' for wavepackets described by PSE solutions (figure~\ref{fig:modelconstruct}b). For the PSE model, we will present both cases with and without coherence decay. As the shock cells are assumed to be axisymmetric and stationary, the frequency and azimuthal dependence are described solely by the properties of the wavepacket. 

The three descriptions of $\bm{q}_t$ have varying levels of complexity. In the simplest description, the perfectly-coherent PSE model only requires a jet mean flow profile and a single parameter to fix the free amplitude of the linear solution. This reduced-order representation should confirm the results of \citet{Tam1987}. As suggested by \citet{Wong2019}, and confirmed in \S~\ref{sec:results}, a linear model is unable to capture certain features of the BBSAN spectrum and a description of the non-linearities in the form of coherence decay is thus imposed on the linear wavepackets. The two simplified cases are compared to the wavepacket obtained from LES data alone, which represents the most accurate description of the current BBSAN model.  

Clearly, a shortcoming of this BBSAN source interpretation \citep{Tam1987, lele2005phased, ray2007sound} is, by construction, the artificial separation of the shock disturbances from the wavepacket. The evolution and dynamics of the wavepacket are assumed independent of the presence of shocks in the jet. Hence, the properties of the educed wavepackets (e.g. convection velocity, phase, amplitude) may differ from those in a shock-containing flow. While there is evidence to suggest that wavepacket dynamics are not affected by weak shocks \citep{edgington2019modulation}, it remains unknown whether this extends to highly underexpanded jets, such as that studied here. Despite PSE having been attempted on a shock-containing base flow \citep{ansaldi2016pse}, that approach is not pursued here, due to the breakdown of the slowly-diverging mean flow assumption in the vicinity of the shocks. 

\section{Nearfield predictions and comparisons} \label{sec:nearfieldcompare}
\subsection{Shock-cell component: Comparison of PIV data and modified P-P model} \label{sec:shockcompare}
\begin{figure}
    \centering
    \includegraphics[width=0.8\linewidth]{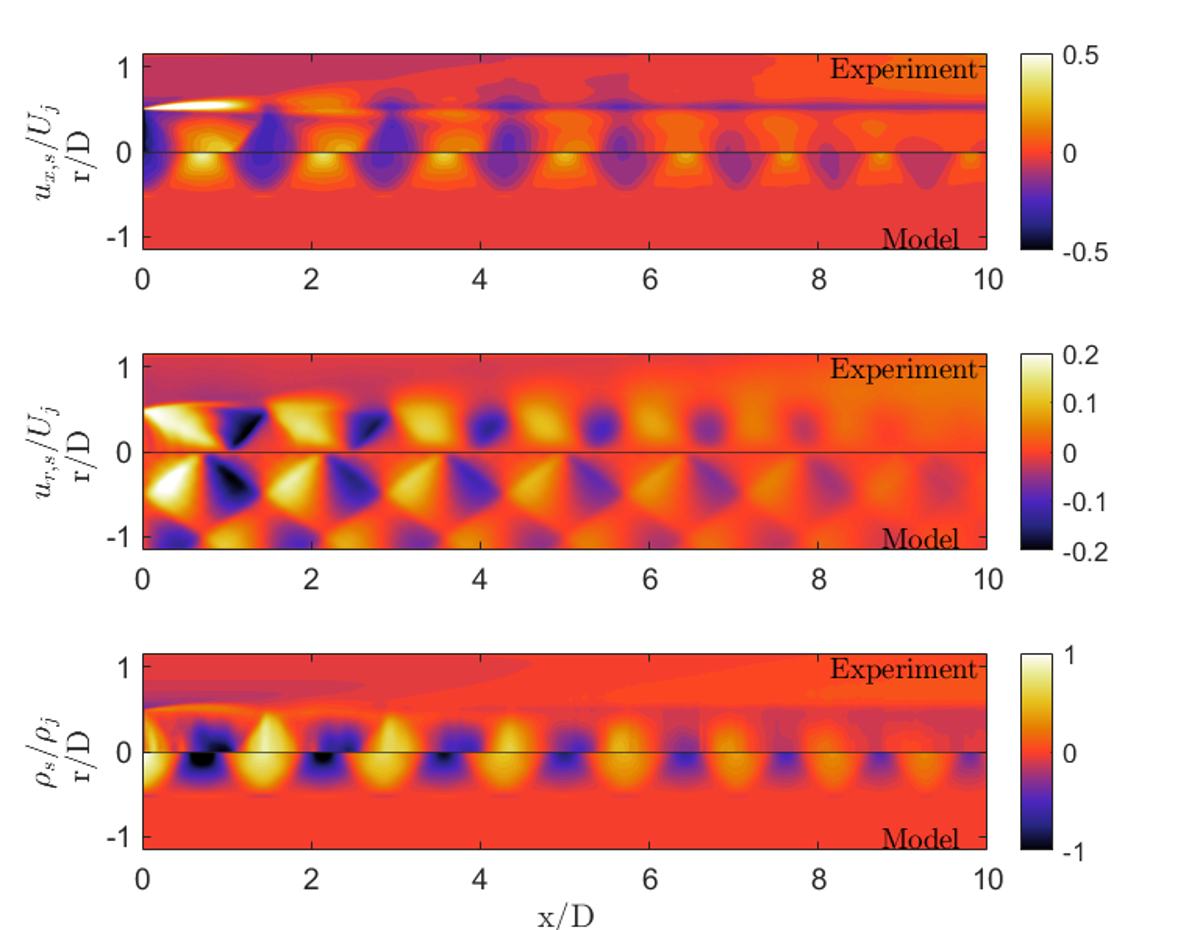}
    \caption{$x-r$ contour plots of flow variables $u_x, u_r, \rho$ from PIV experiments (top-plane) and model (bottom-plane). }
    \label{fig:PPvExpcont}
\end{figure}

Comparisons between the modified P-P model and experimental PIV data for the shock-cell disturbances are shown in figure~\ref{fig:PPvExpcont}. The $x-r$ contour maps show good agreement for each of the flow variables $[u_x, u_r, \rho]_s$ in phase and amplitude. The axial decay in the strength of the shock-cell structure is also well-captured by the model. There is poor agreement in the shear layer region as expected; the model uses a vortex-sheet approximation which is non-physical along the nozzle lip line. It should be noted that the experimental shock-cell disturbances are weak by $x=10D$. Thus, based on this observation and since the equivalent source is the product of $\bm{q}_s$ and $\bm{q}_t$ (equation~(\ref{eq:BBSANTij})), contributions to be BBSAN source for $x>10D$ are negligible. While there remain differences between model and experimental data, figure~\ref{fig:PPvExpcont} illustrates that the salient qualitative features of $\bm{q}_s$ are preserved by the model. Furthermore, we expect these small discrepancies to have minimal impact on the far-field noise as they are dwarfed by other effects, as discussed in appendix~\ref{sec:note}.  

\subsection{Wavepacket component: Comparison of PSE and LES} \label{sec:PSEvLES}
We compare the PSE predictions with the wavepackets extracted from LES data for a selection of frequencies and the first azimuthal mode ($m=0$). The PSE solver used in this study has previously been validated for supersonic flows \citep{kleine2017evaluation}. The aim of this section is not to show in-depth comparisons, but rather to highlight key similarities and differences which may impact the BBSAN source composition. Detailed investigations have previously been carried out by \citet{Cavalieri2013} and \citet{sinha2014wavepacket} for subsonic and supersonic jets respectively. Thus, for brevity, only comparisons for axial velocity fluctuations are shown; a similar degree of agreement is obtained for the remaining components of $\bm{q}_t$. 

It is well known that PSE solutions produce poor agreement with LES data for $St\leq 0.3$, as a weaker KH growth rate becomes comparable with the Orr mechanism induced by non-linear interactions~\citep{tissot2017wave,schmidt2018spectral,pickering2020resolvent}. Discrepancies at low frequencies, however, do not affect the results presented in \S~\ref{sec:results}, since BBSAN dominates at higher frequencies. Hence, comparisons are only shown for $St>0.4$.

For comparison of wavepacket structure, Spectral Proper Orthogonal Decomposition (SPOD) is also performed on the LES data. SPOD decomposes the flow into an orthogonal basis optimally ranked by energy content. The smaller-scale turbulence will be filtered out, highlighting the coherent structures present in the flow. SPOD has been used to show an acceptable degree of fidelity between PSE predictions and SPOD-filtered LES data for the $M=1.5$ jet \citep{rodriguez2013acoustic,sinha2014wavepacket}. For a given azimuthal and frequency, we define the spectral eigenvalue problem \citep{towne2018spectral}

\begin{equation} \label{eq:SPOD}
    \int \mathcal{Q}_{ij}(\bm{x}_1,\bm{x}_2;m,\omega) \bm{\Psi}(\bm{x}_2;m,\omega)d\bm{x}_2 = \lambda(m,\omega)\bm{\Psi}(\bm{x}_1;m,\omega),
\end{equation}
where $\mathcal{Q}_{ij}$ is the cross-spectral density matrix of the flow variable of interest, $\lambda$ and $\bm{\Psi}$ are the eigenvalues and a set of linearly-independent spatial eigenfunctions respectively. Both eigenfunctions and eigenvalues are obtained using the snapshot method described in \citet{towne2018spectral}.

\begin{figure}
    \centering
    \includegraphics[width=1.0\linewidth]{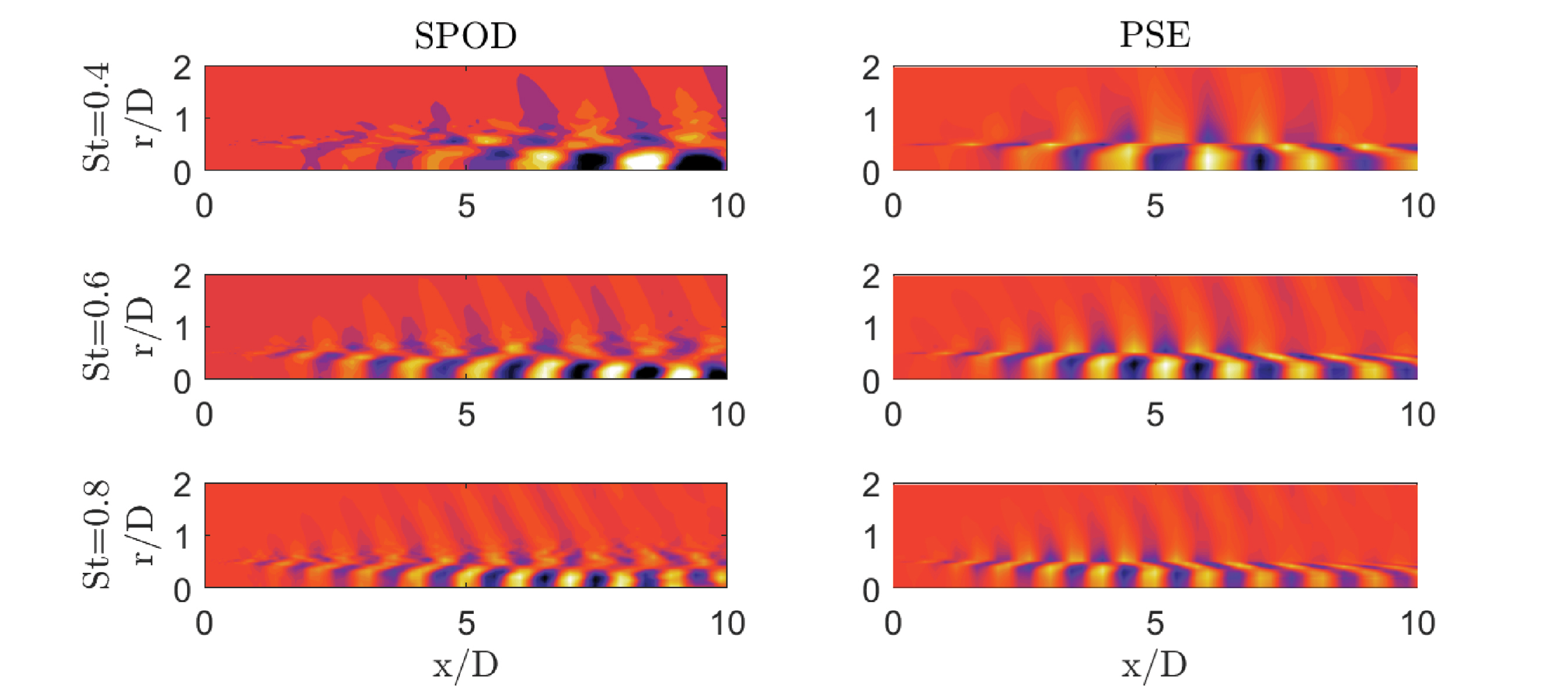}
    \caption{Comparison of real parts of $u_x$ between the extracted wavepacket from the first SPOD mode (left column) and PSE predictions (right column) for $m=0$. Flow quantities are normalised by the ideally-expanded condition.}
    \label{fig:contourm=0}
\end{figure}

Figure~\ref{fig:contourm=0} shows the real component of axial velocity for the axisymmetric mode $m=0$. For each frequency, the PSE solutions (right column) are scaled using the averaged $\mathcal{A}$ constant. The contour maps show the PSE predictions are able to capture both the near-field fluctuations and the propagating Mach wave radiation. As frequency increases, the axial location of the wavepacket peak ($x_0$) shifts upstream and the spatial wavelength decreases. As expected, the mode shapes, wavelength and phase of the PSE and the leading SPOD (left column) fields exhibit good agreement.

\begin{figure} 
    \centering 
    \includegraphics[width=1.0\linewidth]{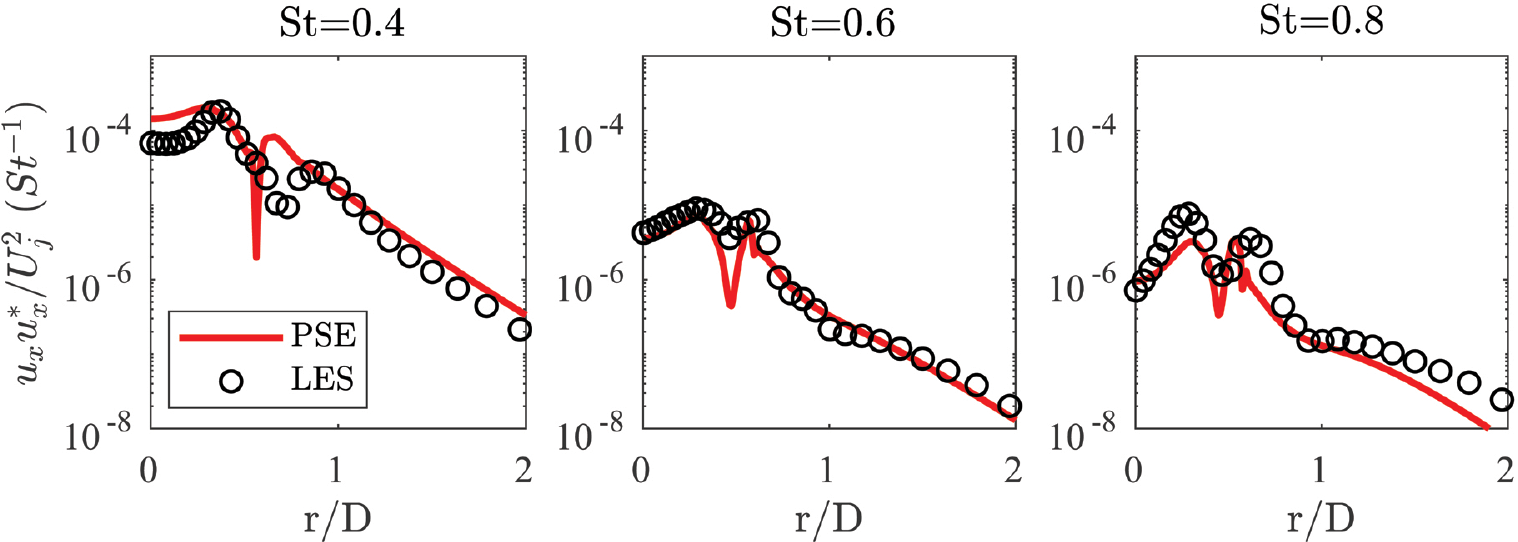}
    \caption{Radial cross-section comparisons of $u_x$ between LES (symbols) and PSE (lines) for $m=0$ at $x=4D$.}
    \label{fig:radial_m00}
\end{figure}

Success in amplitude matching between PSE and LES fields is observed in the radial shapes at the axial station $x=4D$ in figure~\ref{fig:radial_m00}. For the PSE solutions, the drop in amplitude of $u_x$ near the lip line is due to the phase jump either side of the mixing layer in a perfectly-coherent wavepacket \citep{Cavalieri2013}. This is not observed in the LES data due to the jitter of the coherent wavepackets \citep{Cavalieri2013, baqui2015coherence}. By comparing the spatial structure of the shock disturbances shown in figure~\ref{fig:PPvExpcont} with the wavepacket radial profiles, the distributed nature of the BBSAN source is apparent. The wavepacket has non-zero support within the jet potential core, allowing it to interact with the shock-cell structure and generate BBSAN. This will be shown in the source maps presented in \S~\ref{sec:sourcecharacter}.

\begin{figure} 
    \centering 
    \includegraphics[width=0.9\linewidth]{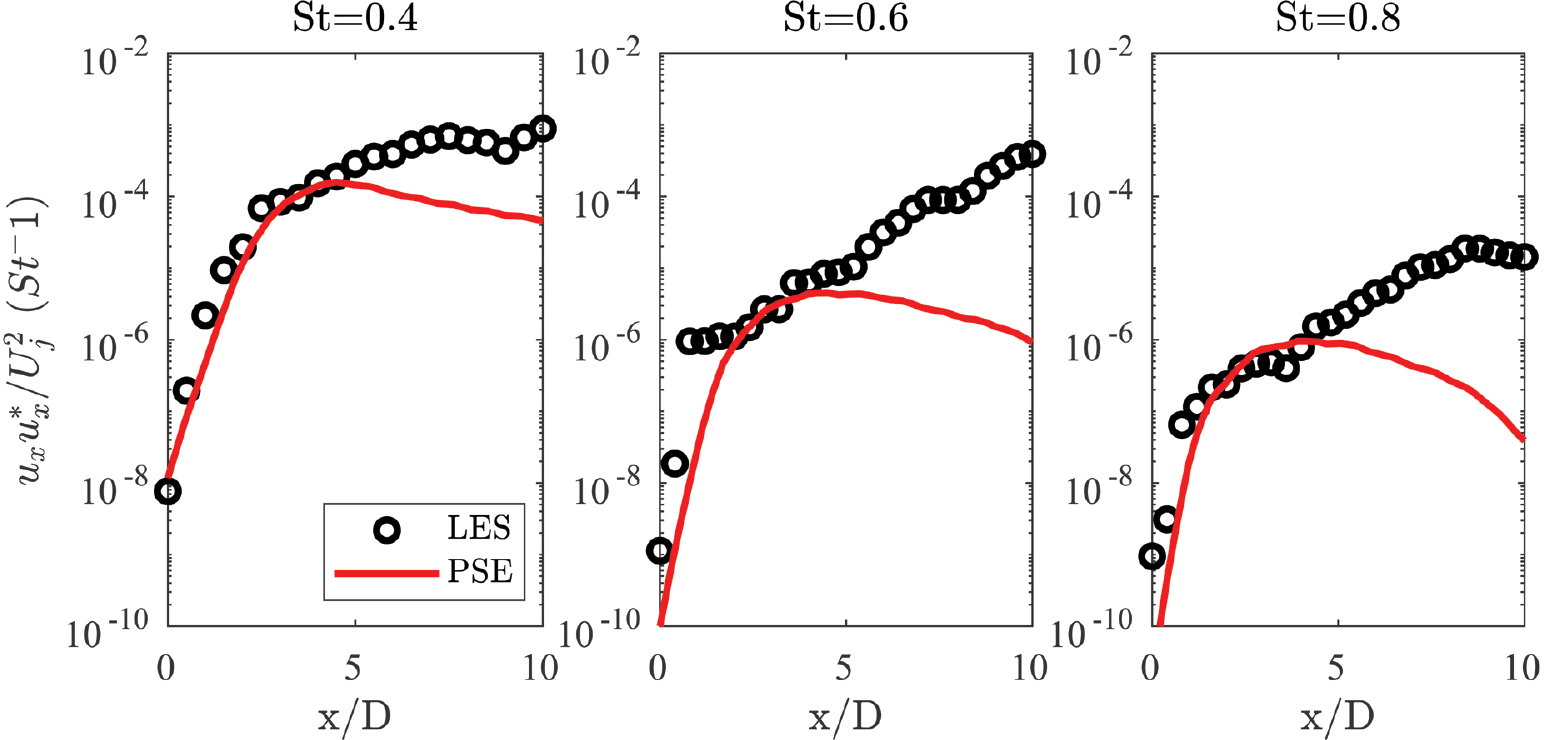}
    \caption{Centreline axial velocity fluctuations from LES (symbols) and PSE (lines) for $m=0$.}
    \label{fig:centreline_m00}
\end{figure}

The centreline axial velocity fluctuations in figure~\ref{fig:centreline_m00} increase in energy by approximately four orders of magnitude between the nozzle exit and the location of the peak value ($x_0 \approx 5D$). This amplification is also observed in hot-wire measurements in subsonic jets \citep{Cavalieri2013}. We also observe that, relative to the LES data, PSE underestimates amplitudes in the downstream portion of the jet ($x>6D$). This well-known inconsistency has previously been attributed to the dominance of non-linear effects, and fluctuations that are uncorrelated with the extracted wavepackets \citep{suzuki2006instability,gudmundsson2011instability,Cavalieri2013}. Since shock fluctuations remain significant past $x=5D$ (figure~\ref{fig:PPvExpcont}), the discrepancy in turbulent intensity may lead to differences in BBSAN prediction between the LES and PSE model. A more detailed discussion of this issue can be found in \S~\ref{sec:sourcecharacter}. 

\begin{figure} 
    \centering 
    \includegraphics[width=0.9\linewidth]{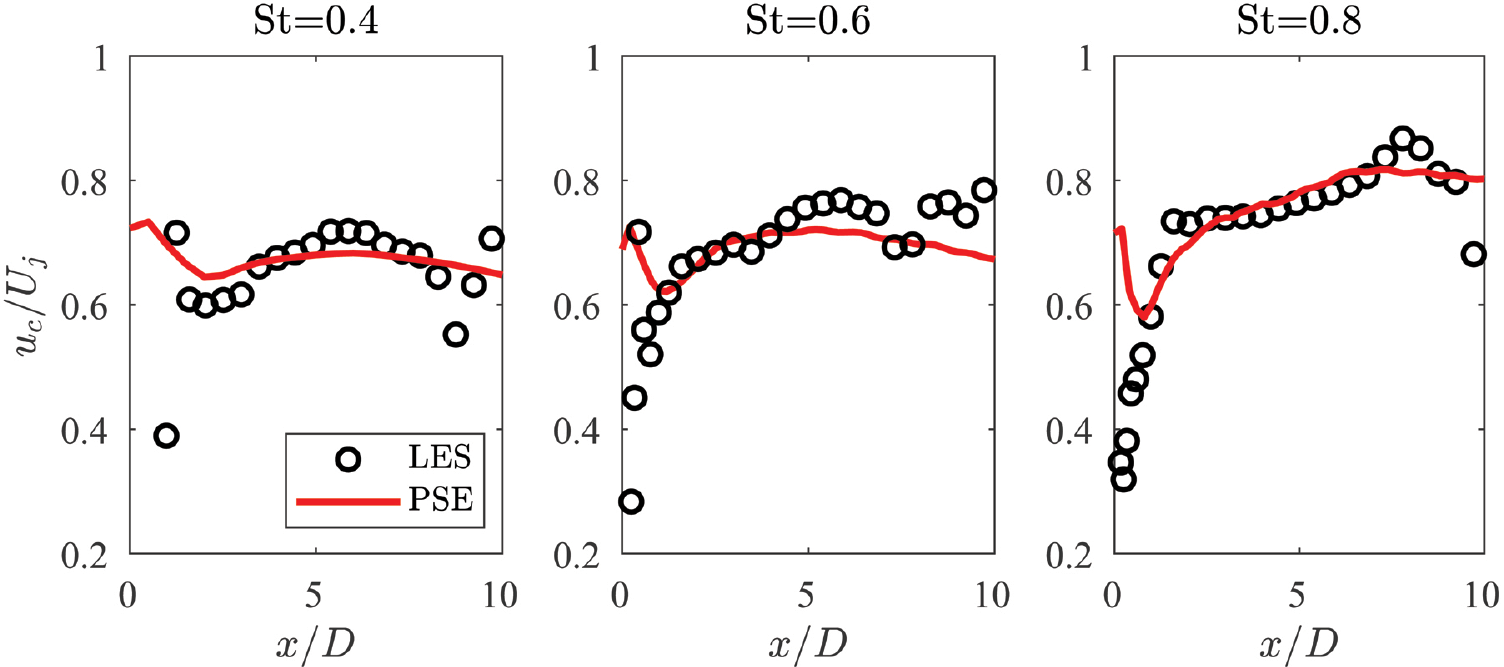}
    \caption{Convection velocity as a function of axial position for $m=0$.}
    \label{fig:uc_m00}
\end{figure}

Lastly, from equation~(\ref{eq:HBFtam}), it is evident that the BBSAN peak frequency strongly depends on the convection velocity of the large-scale structures. The convection velocity is related to the hydrodynamic wavenumber $k_h$, which is extracted from the PSE solution as the real component of the eigenvalue $\alpha_{m,\omega}$ 

\begin{equation} \label{eq:ucPSE}
    u_c(x_1) = \frac{\omega}{k_h} = \frac{2\pi St}{Re(\alpha_{m,\omega}(x_1))} .
\end{equation}
For the LES case, $u_c$ can be computed using the argument $\phi$ of the CSD \citep{maia2019two},

\begin{equation} \label{eq:ucLES}
    u_c(x_1) = \frac{\omega}{k_h} = \omega \left( \frac{\partial \phi}{\partial x_2}\right)^{-1} .
\end{equation}
Figure~\ref{fig:uc_m00} shows the extracted $m=0$ phase velocities for PSE predictions (equation~(\ref{eq:ucPSE})) and the LES results (equation~(\ref{eq:ucLES})). Over a range of frequencies, $u_c$ is estimated as $\approx 0.7-0.8U_j$ over much of the flow domain. Despite disagreements within the first diameter, agreement improves further downstream. This result suggests that both PSE and LES-based sources should predict comparable BBSAN peak frequencies according to equation~(\ref{eq:HBFtam}). 

We have shown that many of the wavepacket features extracted from LES are reproducible with PSE. In line with previous studies \citep{rodriguez2013acoustic, sinha2014wavepacket, sasaki2017high}, good agreement is also observed at higher azimuthal wavenumbers. We reiterate that our goal is not to find optimal agreement between PSE model and LES data, but rather, to compute an appropriate scaling parameter for the indeterminant PSE amplitude. 

\section{Far-field acoustic spectra and comparisons with experiment}\label{sec:results}
Far-field acoustic predictions based on the BBSAN source models are examined in comparison with the experimental far-field noise measurements detailed in section \S~\ref{sec:acousticdata}. There are some points to be highlighted in the presentation of these results. Firstly, we reiterate that, apart from the modifications to the P-P shock-cell model and scaling of the PSE to the LES data, the source is entirely built from flow information alone. The shock-cell representation used for both PSE and LES-based models is identical.

As shown in figure~\ref{fig:coord}, the polar angle $\theta$ is nominally taken from the downstream jet axis. Since the acoustic measurements are taken along a cylindrical surface at a moderate distance of $R=11D$ from the jet centreline, the origin of the polar angle is moved to $X_o = 5D$ instead of the nozzle exit. This modification enables comparison with directivity results from other far-field jet databases in literature, where microphones are placed much further from the jet, and also provides a small correction in predictions of peak frequency which is consistent with equation~(\ref{eq:HBFtam}). 

After computing the far-field PSD from equation~(\ref{eq:psd2}), the sound pressure level (SPL) is defined by
\begin{equation}
    SPL = 10\log_{10}\left(\frac{\langle pp^*\rangle}{p^2_{ref}}\right) 
\end{equation}
where $p_{ref} = 20\mu Pa$ and SPL is in units of dB/St. 

\subsection{Directivity contour maps} \label{sec:contour}
\begin{figure}
    \centering
    \begin{subfigure}{1.0\textwidth}
    \hspace*{-.5cm} 
    \includegraphics[width=1.0\linewidth]{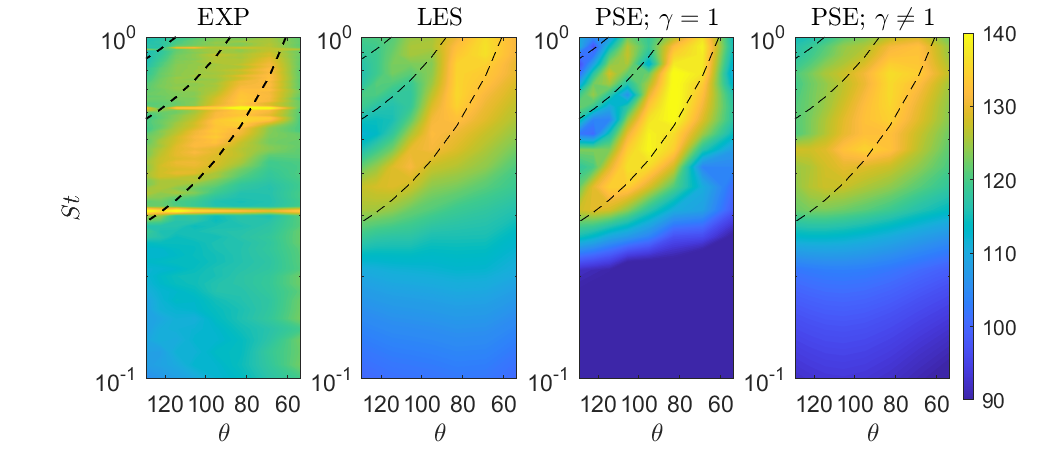}
    \caption{m=0.}
    \end{subfigure}
    \begin{subfigure}{1.0\textwidth}
    \hspace*{-.5cm} 
    \includegraphics[width=1.0\linewidth]{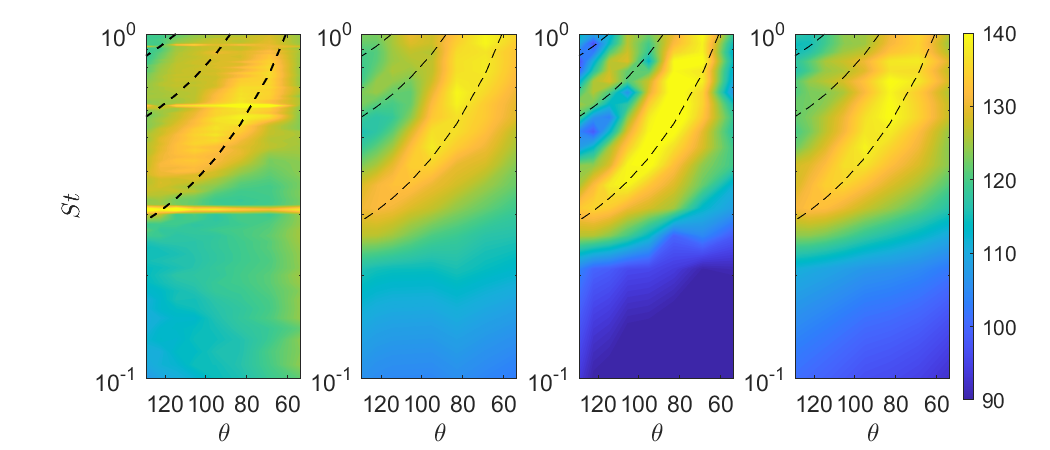}
    \caption{m=1.}
    \end{subfigure}
    \begin{subfigure}{1.0\textwidth}
    \hspace*{-.5cm} 
    \includegraphics[width=1.0\linewidth]{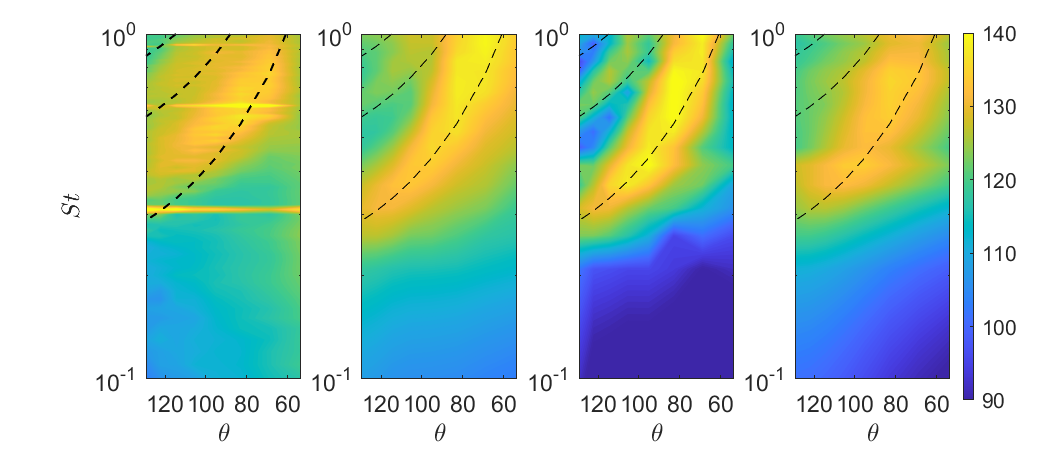}
    \caption{m=2.}
    \end{subfigure}
    \caption{$St-\theta$ directivity contour maps of sound pressure level spectra at $R=11D$. Contours are in dB/St. }
    \label{fig:contourBBSAN}
\end{figure}

To observe the spectra and directivity trends of BBSAN, we first present $St-\theta$ contour maps in figure~\ref{fig:contourBBSAN}, from experimental data and model predictions. Unlike \citet{Tam1987} and \citet{ray2007sound}, who compared predictions to the full acoustic signal, we retain the dependence on azimuthal wavenumber and show results for the first three modes ($m=0$, 1 and 2). To highlight the theoretical BBSAN peak locations, peak frequencies computed using equation~(\ref{eq:HBFtam}) are also indicated as dashed lines for the first three shock-cell modes ($n=1$, 2 and 3), where we have assumed the convection velocity to be $u_c=0.7U_j$. 

As expected in the first column of figure~\ref{fig:contourBBSAN}, the experimentally measured BBSAN lobe is visible for $St>0.4$ between $65^{\circ}<\theta<120^{\circ}$, and peak frequency increases as observer position moves downstream. Screech peaks are clearly discernible as discrete frequencies, with the fundamental located at $St=0.31$. The BBSAN primary lobe agrees largely with the theoretical peak frequency prediction at sideline and downstream positions, though some discrepancy develops at more upstream angles ($\theta>110^{\circ}$). This could be due to the measurements not being performed in the `true' far-field, or may arise from the variation in convection velocity as a function of frequency. The frequency of the second shock-cell mode ($n=2$) peak is consistently higher than theory, which may arise from the mismatch in Mach numbers (and hence shock-cell spacing), between the PIV and acoustic databases (see figure~\ref{fig:Mj}).

To accompany the measured acoustics, figure~\ref{fig:contourBBSAN} provides predictions based on the BBSAN source models of \S~\ref{sec:sourceconstruct}. We present three models for the reconstructed BBSAN source, each with a different description of $\bm{q}_t$. The LES model is presented in the second column of figure~\ref{fig:contourBBSAN}, while those described by PSE solutions with unit coherence or with coherence decay are shown in columns three and four respectively. We note that discrete peaks do not feature in either of the LES or PSE model predictions, as the non-linear screech mechanism is not modelled. In addition, significant underprediction occurs at low frequencies ($St<0.4$), as expected; the source term in equation~(\ref{eq:BBSANTij}) includes only the high-frequency BBSAN component.

Far-field noise predictions using the LES model exhibit fair agreement with measured data across a wide frequency and directivity range. The best agreement is in the sideline direction for both amplitude and peak frequency predictions; the LES model matches the experimental measurements to within $\pm$ 2dB/St. The model follows the theoretical BBSAN peak from equation~(\ref{eq:HBFtam}), even at upstream angles where the peak half-width narrows. This is unsurprising since equation~(\ref{eq:HBFtam}) assumes that BBSAN is produced by the interaction of an instability wave with the stationary shock-cell structure, with the resulting difference waves effectively behaving as the source of the far-field noise. In addition, the convection velocity of the extracted LES wavepacket (figure~\ref{fig:uc_m00}) is approximately $0.7U_j$. The narrowing of the BBSAN lobe at upstream angles is also observed in the acoustic measurements of \citet{norum1982measurements}.  

Nevertheless, there remain key differences between the LES model and measurements. At slightly downstream angles, overprediction occurs at high frequencies ($St\approx 1$). The overprediction in sound amplitude results in the BBSAN lobe being broader in directivity than the experimental spectra for all three azimuthal modes. The mismatch could be related to the simplification of the Lighthill stress tensor $T_{ij}$, where cancellation between different components is known to occur \citep{Freund2001}. \citet{bodony2008low} found, for a $M_j=2.0$ ideally-expanded jet, that using only the momentum term ($\rho u_iu_j$) overpredicts the sound amplitude by over 20dB/St at high frequencies. Since we retain the momentum term alone (\ref{eq:Tij}), cancellation effects due to entropic and higher-order terms of the equivalent BBSAN source are not accounted for. The definition and simplicity of the present model prevents an investigation into the relevance of this potential phenomenon. 

Despite the simplicity, predictions based on the reduced-order PSE model are also encouraging. The primary BBSAN lobe is well-predicted and has similar trends to that of the LES model. This indicates that the preposition of \citet{Tam1982}, that BBSAN is generated as a result of the interaction between the quasi-periodic shocks and large-scale turbulent structures, is indeed well-founded. Agreement in both peak frequency and amplitude in the present results further substantiates the applicability of the interpretation of \citet{Tam1982}. For upstream angles, the assumption of perfectly-coherent wavepackets is found to result in overprediction of peak intensity, as well as marked dips in the spectra between primary and secondary shock-cell mode signatures. When coherence decay is incorporated, however, the directivity map is smoothed and the dips are reduced. This effect was reported by \citet{Wong2019} for a simple equivalent line-source model. Directivity changes occur as the source energy is spread in wavenumber space between shock-cell modes. By comparing the predictions from both PSE and LES-based models with experimental measurements, it is clear that a linear wavepacket model requires modification to account for non-linearities (e.g. wavepacket jitter) in order to successfully predict BBSAN amplitude. The effects of coherence decay are examined in \S~\ref{sec:spectra}.

\subsection{Far-field noise spectra} \label{sec:spectra}
\begin{figure}
    \centering
    \includegraphics[width=0.9\linewidth]{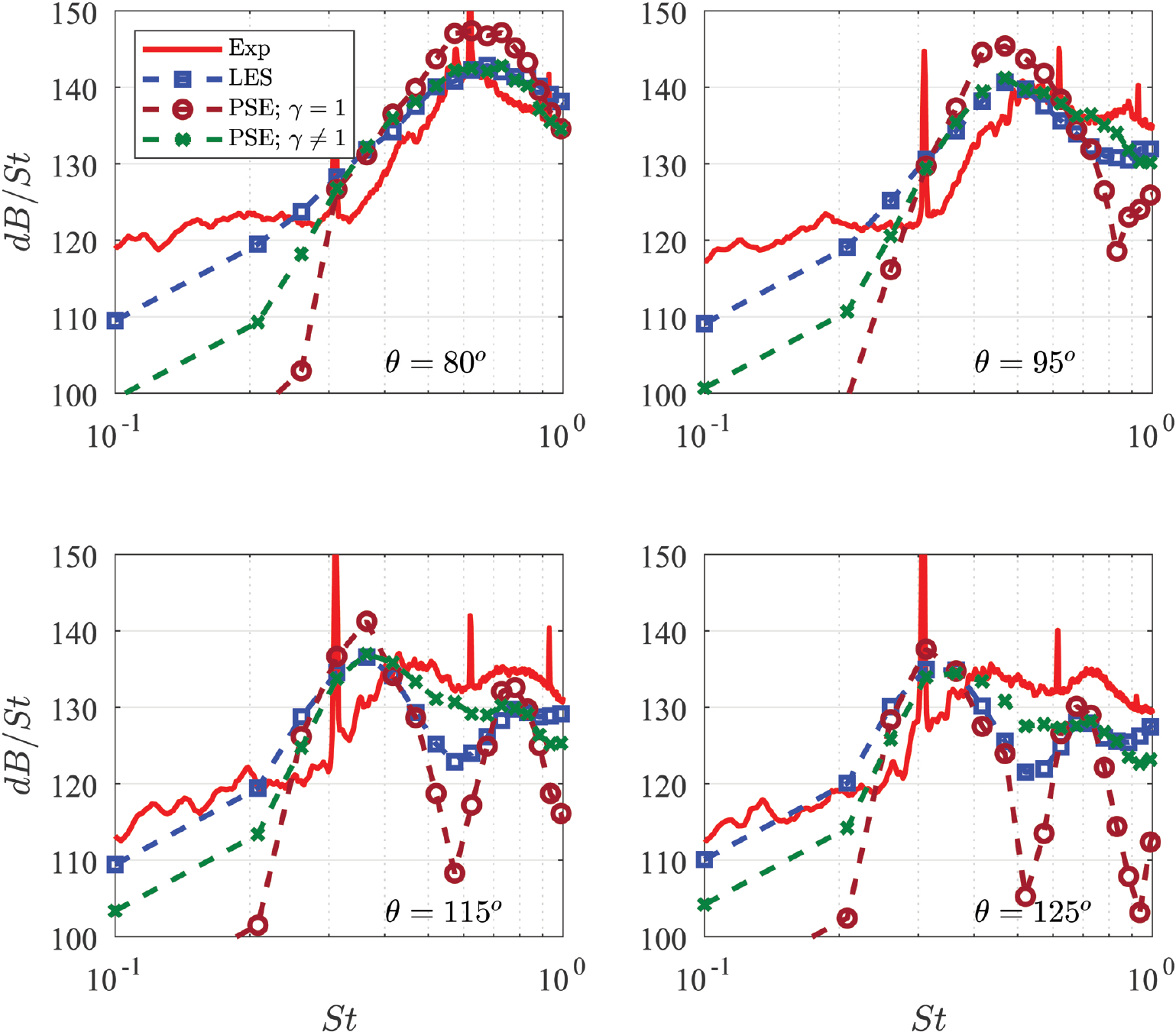}
    \caption{Comparison of acoustic spectra for total measured signal and reconstructed model using the first three azimuthal modes $m=0,1,2$.}
    \label{fig:spectram0-2}
\end{figure}

Before showing azimuthally-decomposed spectra, the total measured sound-field is presented along with reconstructed model predictions using the first three azimuthal modes in figure~\ref{fig:spectram0-2} at different polar angles. For each observer position, predictions from both the LES (blue squares) and PSE models are shown, along with the full (solid red) acoustic spectra. The PSE predictions are further distinguished by either unit coherence (maroon circles) or coherence decay (green crosses). As shown in the contour directivity plots in figure~\ref{fig:contourBBSAN}, the models miss the peak BBSAN frequency at upstream angles. Nevertheless, excellent agreement in peak amplitude is observed ($\pm$ 2dB/St) for the primary ($n=1$) peak across the directivity range. Even with a small number of inputs, the simplified PSE model with perfect coherence performs particularly well in capturing peak amplitudes, though large dips are observed as either the polar angle or frequency increases. There is less success in predicting the secondary lobe ($n=2$) due to its increased azimuthal modal complexity, requiring 4-5 modes to reconstruct the total signal \citep{wong2020azimuthal}.

\begin{figure}
    \centering
    \includegraphics[width=0.9\linewidth]{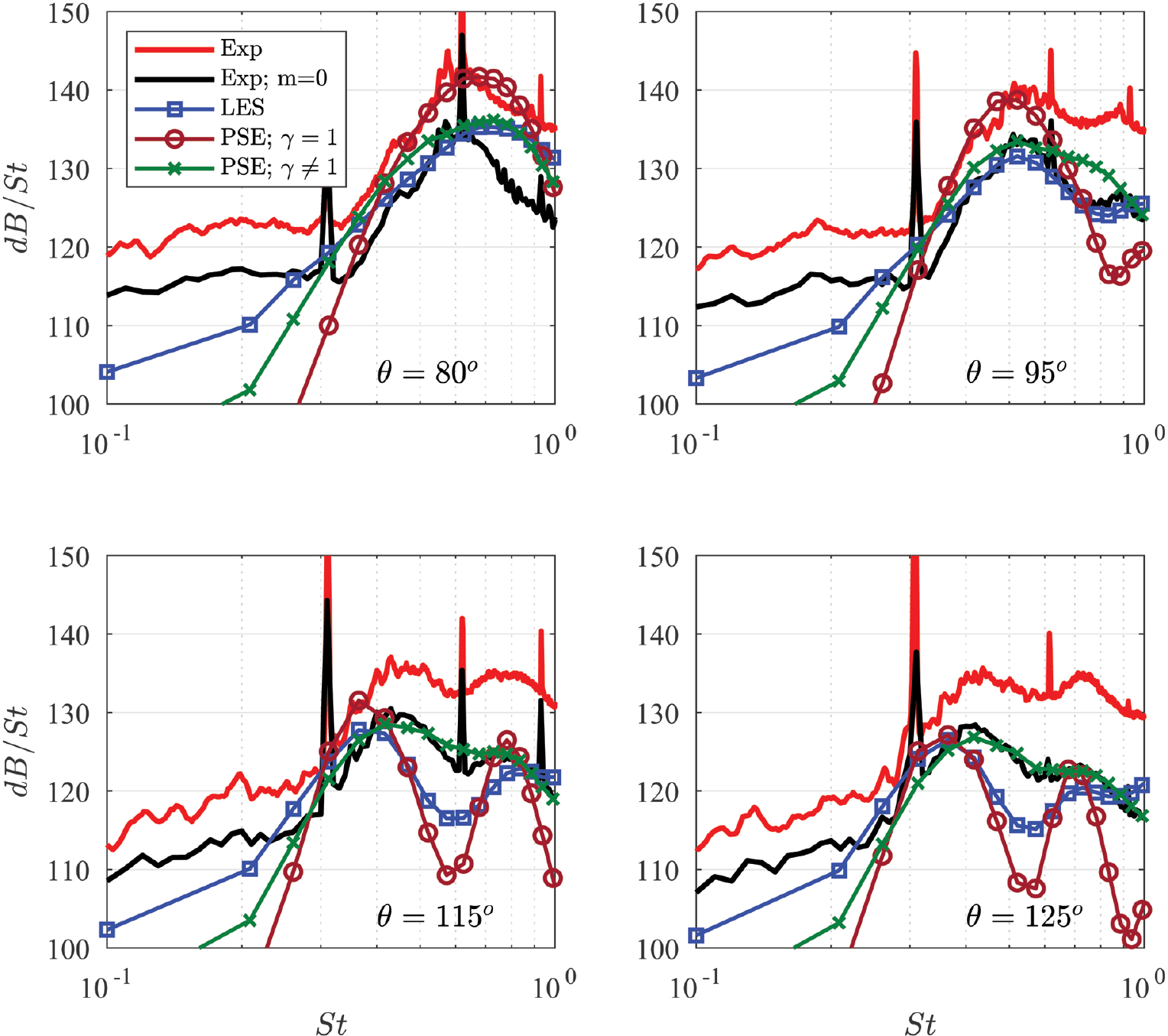}
    \caption{Comparison of acoustic spectra for azimuthal mode $m=0$.}
    \label{fig:spectram00}
\end{figure}

\begin{figure}
    \centering
    \includegraphics[width=0.9\linewidth]{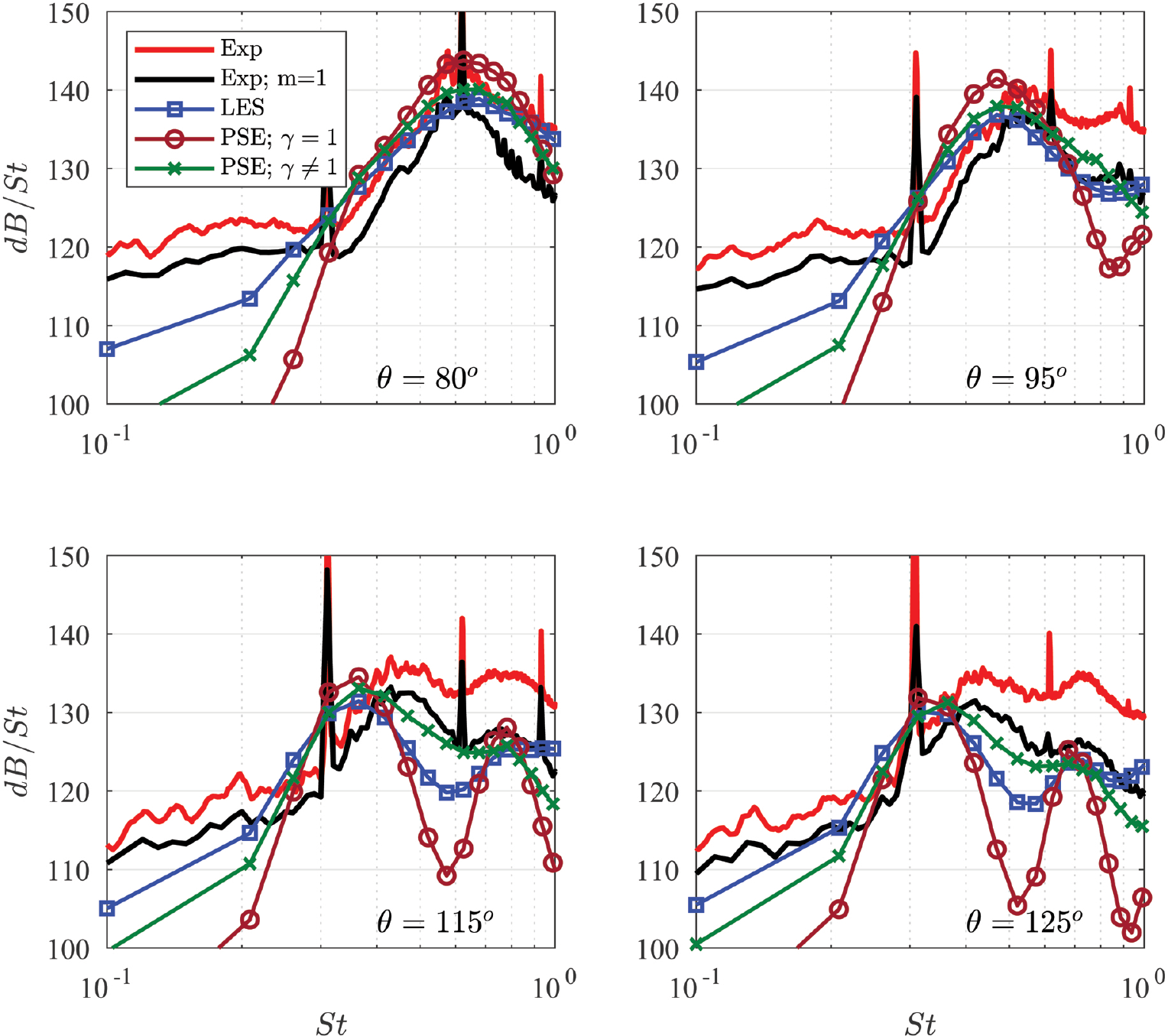}
    \caption{Comparison of acoustic spectra for azimuthal mode $m=1$.}
    \label{fig:spectram01}
\end{figure}

\begin{figure}
    \centering
    \includegraphics[width=0.9\linewidth]{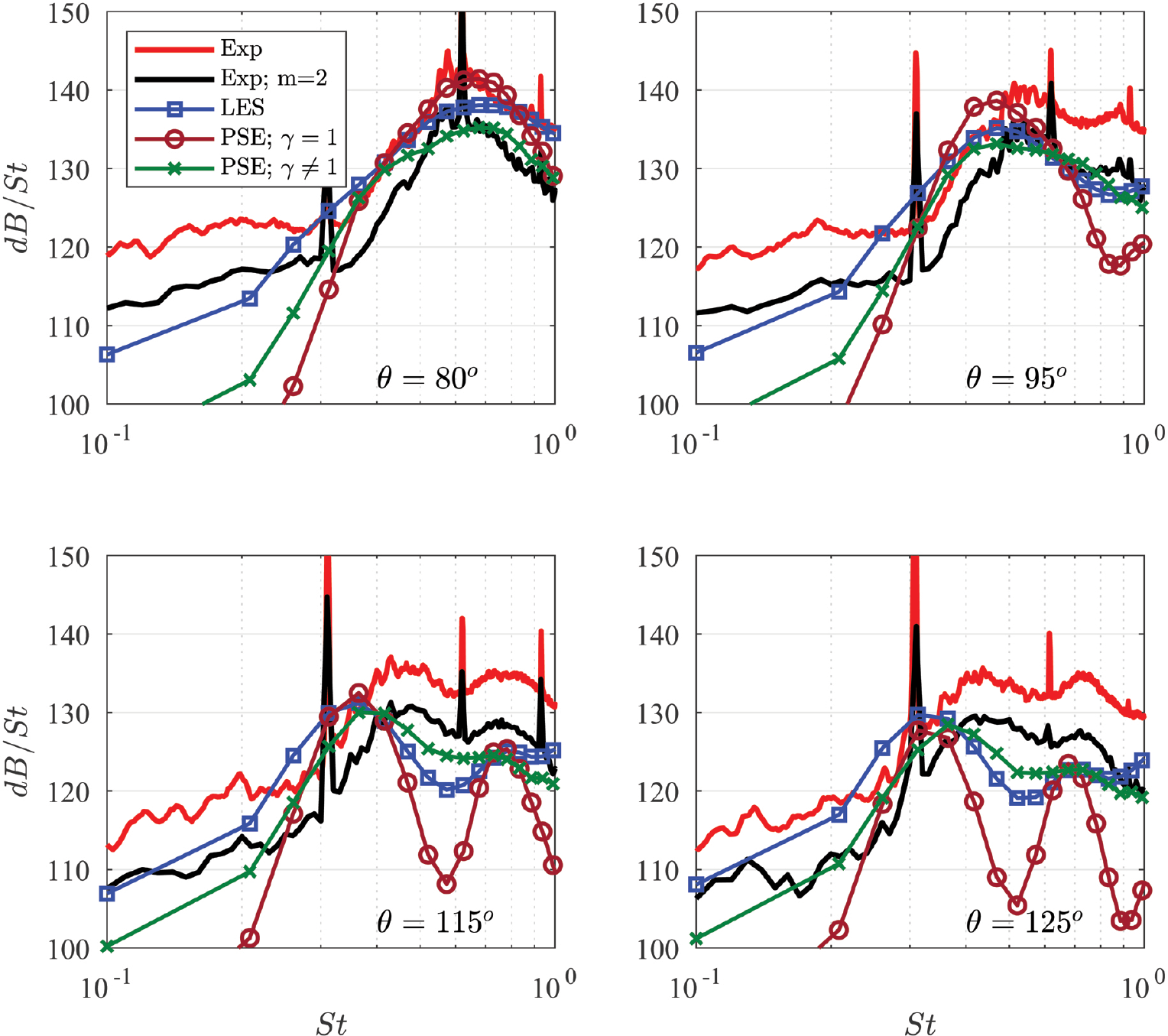}
    \caption{Comparison of acoustic spectra for azimuthal mode $m=2$.}
    \label{fig:spectram02}
\end{figure}

To explore the similarities and differences between experimental and model spectra in further detail, figures~\ref{fig:spectram00}-\ref{fig:spectram02} provide spectra for each of the azimuthal modes. In addition to the total signal, azimuthally-decomposed data (solid black) is shown. In terms of peak frequency and amplitude, we observe fair agreement between models and experiment for both the primary and secondary BBSAN peaks. Peak amplitudes are within $\pm$ 2dB/St accuracy and predicted peak half-width is most faithful to the measured spectra in the sideline direction ($\theta=95^{\circ}$). 

Previous studies have compared stability-based BBSAN models to the total acoustic signal (similar to figure~\ref{fig:spectram0-2}). Ambiguity in amplitude of model predictions has led to the azimuthal dependence being dropped; \citet{ray2007sound} assumed a `white noise' spectrum while \citet{Tam1987} assumed the equivalent source to be solely axisymmetric. The spectra of the equivalent source models are then fitted to experimental acoustic data. The ill-posed nature of such `outside-in' approaches may lead to the deduction of source parameters not observed in the jet. Indeed, the azimuthally-decomposed acoustic spectra provided in figures~\ref{fig:spectram00}-\ref{fig:spectram02} and the recent measurements performed by \citet{wong2020azimuthal} indicate that these assumptions are invalid. For instance, the roll-off at high frequencies of individual azimuthal modes is steeper than the total signal \citep[c.f.][]{ray2007sound}, and the spectral shape of each azimuthal mode is not identical \citep[c.f.][]{Tam1987}.

Using a direct `inside-out' approach, inconsistencies in previous BBSAN amplitude predictions are now nullified. Examination of each individual azimuthal mode suggests that the proposed model can correctly capture the important flow dynamics related to BBSAN. Along with the findings from \citet{Wong2019}, the results also offer a convincing explanation for the `missing sound' at high frequencies, as observed by both \citet{suzuki2016wave} and \citet{ray2007sound} at upstream angles. It is clear that the secondary BBSAN peak is due to the interaction of the wavepacket with the second shock-cell mode which was not accounted for in either study.

As alluded to in \S~\ref{sec:contour}, there are regions where the models perform poorly. At upstream angles ($\theta=115^{\circ}$ and $125^{\circ}$), while the agreement in peak amplitude is within $\pm$2dB/St, peak frequency is underpredicted. At slightly downstream positions ($\theta=80^{\circ}$), the predicted half-width of the primary BBSAN peak is larger than measured. As well as the overprediction at high frequencies, the second harmonic of the screech tone coinciding with the BBSAN peak may explain why the models predict higher peak frequencies ($St_p \approx 0.6$) than the experiment ($St_p \approx 0.55$). The presence of screech is known to attenuate the axial extent of downstream shock cells \citep{andre2013broadband}. Currently, this cannot be verified as flow measurements are not available to supplement the acoustic database. 

We turn our focus to comparing the efficacy of our models. With minimal inputs, the reduced-order model using a perfectly coherent ($\gamma=1$) wavepacket source does a respectable job in predicting the primary and secondary BBSAN peaks ($n=1,2$). This is a confirmation of the modelling approach first proposed by \citet{Tam1982}; BBSAN is generated by the interaction between large-scale coherent structures and the shock-cell system. In terms of peak noise in the far-field, it is clear that second-order statistics of the flow are unimportant. The ability for a simple model to capture both amplitude and peak frequency renders it a promising candidate for future predictive schemes. 

Away from the peaks, however, the linear wavepacket source presents some drawbacks. In particular, the `dips' mentioned previously are evident; the discrepancy is more severe at upstream angles, reaching up to 20dB/St less than the measured spectra. The amplitude prediction of the primary peak also becomes questionable over downstream angles (by up to 10dB/St). Agreement in SPL is recovered with the inclusion of two-point coherence information. The improvement was predicted using a model line-source problem \citep{Wong2019}, which included coherence information to represent the jittering of wavepackets due to the action of background turbulence \citep{zhang2014just, tissot2017jfm}. Together with the LES model, which is the most complete representation of the source CSD, figures~\ref{fig:spectram00}-\ref{fig:spectram02} demonstrate the appropriateness the proposed BBSAN modelling framework.  

The dips in figures~\ref{fig:spectram00}-\ref{fig:spectram02} are similar to those observed by \citet{Tam1987}, attributed in that study to shock-cell unsteadiness due to interaction with turbulence. It was suggested that the fluctuating motion of the shocks could lead to further peak broadening, with the maximum shock-cell unsteadiness located near the end of the potential core. A quantitative measure for shock-cell unsteadiness was not available at the time and an empirical adjustment to the source structure was made to account for this effect. We show, however, that in fact most of the broadening is instead attributable to wavepacket jitter; non-linear effects acting on the linear wavepackets are educed from the LES data as coherence decay and imposed onto the PSE model. While a large portion of the `missing sound' can be attributed to wavepacket jitter (up to 15dB/St), the dips are not entirely eliminated in the LES model spectra (e.g. $St=0.6$ for $\theta=115^{\circ}$). In reality, the shock structure is unsteady and this phenomenon is not captured by the model (\S~\ref{sec:summaryconstruct}) since the shocks are modelled as zero-frequency waves. The application of $\bm{q}_s$ and $\bm{q}_t$ as distinct variables in our model further restricts the ability to describe how turbulence affects the shocks, and vice-versa. In addition, apart from unsteadiness due to large-scale structures \citep{Tam1987}, periodic shock oscillations in a screeching jet (such as the one used presently) could be attributed to the passage of upstream-travelling acoustic waves \citep{panda1998shock, edgington2018upstream} or coupling between the shock cells. Due to the current modelling framework, the effects of shock unsteadiness on BBSAN remains unknown.

Based on the above observations, we might hypothesise that the prevailing discrepancies evident in figures~\ref{fig:spectram00}-\ref{fig:spectram02} indicate that both wavepacket jitter (modelled as coherence decay deduced from an ideally-expanded jet) and shock unsteadiness are essential to the composition of an equivalent BBSAN source. Another possibility is that the measure of coherence in a shock-containing jet differs nontrivially to that of an ideally-expanded jet. Investigation into such a coupling between wavepacket dynamics and the shock structure is outside the scope of this study, but ought to be considered in future work.

At upstream angles ($\theta=115^{\circ}$ and $125^{\circ}$), we also observe that the PSE model with coherence decay (green crosses) gives more favourable predictions than the LES model when compared to the measured spectra. This is somewhat unexpected since for the LES model, flow variables are directly substituted into the source CSD, while the PSE solution only provides the statistical wavepacket. From equation~(\ref{eq:Gamma}), an adequate description of the original acoustic source requires matching of not only average amplitude and phases of wavepackets (provided by the PSE), but also a correct description of the two-point coherence function. A mismatch in the description of any one of these physical traits will translate into disagreement in the predicted acoustic field. We explore this inconsistency in \S~\ref{sec:sourcecharacter} by inspecting the reconstructed BBSAN sources. 

\section{Source term characteristics} \label{sec:sourcecharacter}
This sections aims to highlight the differences between the reconstructed sources using the various descriptions for $\bm{q}_t$ (LES, PSE with and without coherence decay). For brevity, we will only show the $S_{11}$ component for the $m=0$ azimuthal mode at frequencies $St=0.6$ and $0.8$. The other source term components and azimuthal modes display similar behaviour. 

\begin{figure}
    \centering
    \begin{subfigure}{0.8\textwidth}
    \includegraphics[width=1.0\linewidth]{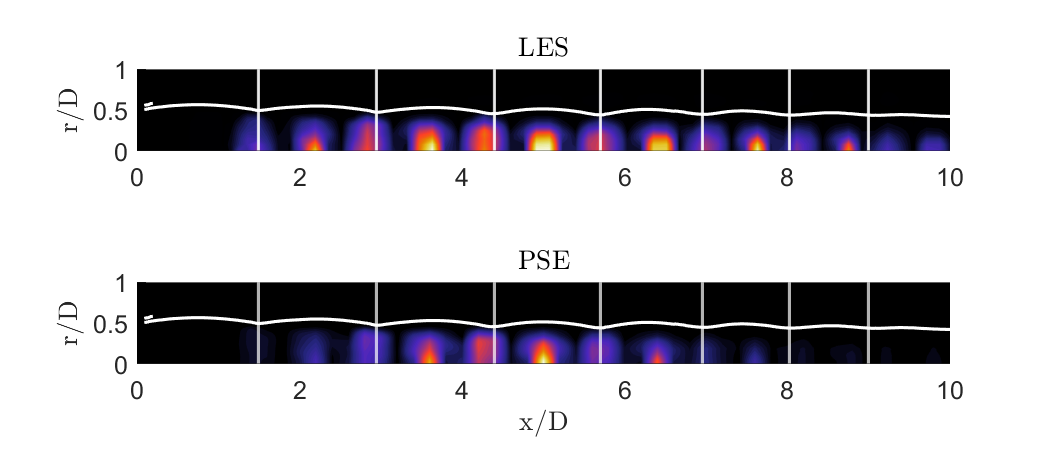}
    \caption{$St=0.6$}
    \label{fig:sourcemapst06}
    \end{subfigure}
        \begin{subfigure}{0.8\textwidth}
    \includegraphics[width=1.0\linewidth,]{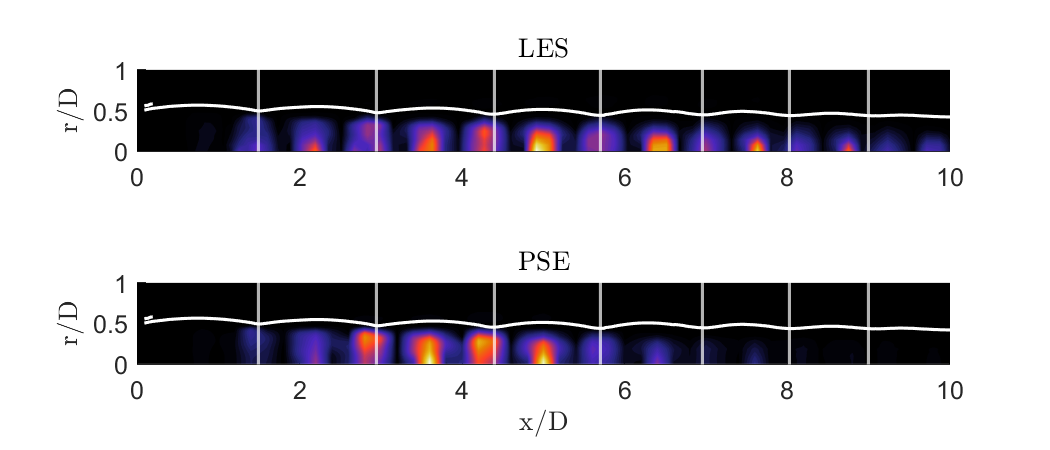}
    \caption{$St=0.8$}
    \label{fig:sourcemapst08}
    \end{subfigure}
    \caption{$x-r$ contour maps of the reconstructed BBSAN source PSD. Intensity levels are normalised by the maximum value.} \label{fig:sourcemap}
\end{figure}

Figure~\ref{fig:sourcemap} shows the reconstructed BBSAN sources for both the LES and PSE cases. At each radial station, the PSD of the source in equation~\ref{eq:sij} is plotted by setting $x_1=x_2$. The sources of the two PSE cases ($\gamma=1$ and $\gamma\neq1$) are identical, since the inclusion of coherence decay does not affect the PSD. To aid in visualisation of the shock positions, the sonic line of the jet plume and the shock-reflection points from the PIV data are shown. Contour levels are normalised by the maximum level. Unlike subsonic jets \citep{maia2019two}, we do not observe a smooth asymmetric Gaussian envelope. Due to the interaction with the shocks, the source is semi-distributed in both axial and radial directions. For each shock cell, there are two source locations; just upstream of the compression-wave focus and before the shock reflection points. Unlike the source maps of \citet{kalyan2017broad} and \citet{tan2018equivalent} which are focused on the sonic line in the shear layer, the source exists inside the jet plume. The present distributions are supported by other models \citep{ray2007sound, shen2020extraction} and also experimental measurements \citep{savarese2013experimental}. Source intensity is apparent between $2D\leq x \leq 8D$ downstream, and most intense between the third and fifth shock cells. This is slightly upstream compared to those measured by \citet{norum1980location} and \citet{seiner1984acoustic} for underexpanded jets operating at similar conditions. As frequency increases, the wavepacket contracts (figure~\ref{fig:contourm=0}) and hence the source shifts towards the nozzle, in line with previous modelling efforts \citep{ray2007sound, suzuki2016wave, patel2019statistical}.

Evidently, the LES description has source intensity extending past $x=8D$ while the PSE models do not. This is due to the differences between the LES and PSE description of the wavepacket; the PSE solution is unable to capture the downstream incoherent fluctuations as discussed in \S~\ref{sec:PSEvLES}, and as shown in figure~\ref{fig:centreline_m00}. This observation may explain the difference in far-field predictions between the LES and PSE with coherence decay ($\gamma \neq 1$) case. As mentioned in \S~\ref{sec:spectra} and by \citet{cavalieri2014coherence}, agreement between the original and statistical source requires the coherence, in addition to both average amplitude and phase, of wavepackets to be the same. Since two-point coherence information imposed on the PSE model is extracted directly from LES data, any difference in the far-field will arise from a mismatch in the average wavepacket envelope shape.

\begin{figure}
    \centering
    \includegraphics[width=1.0\linewidth]{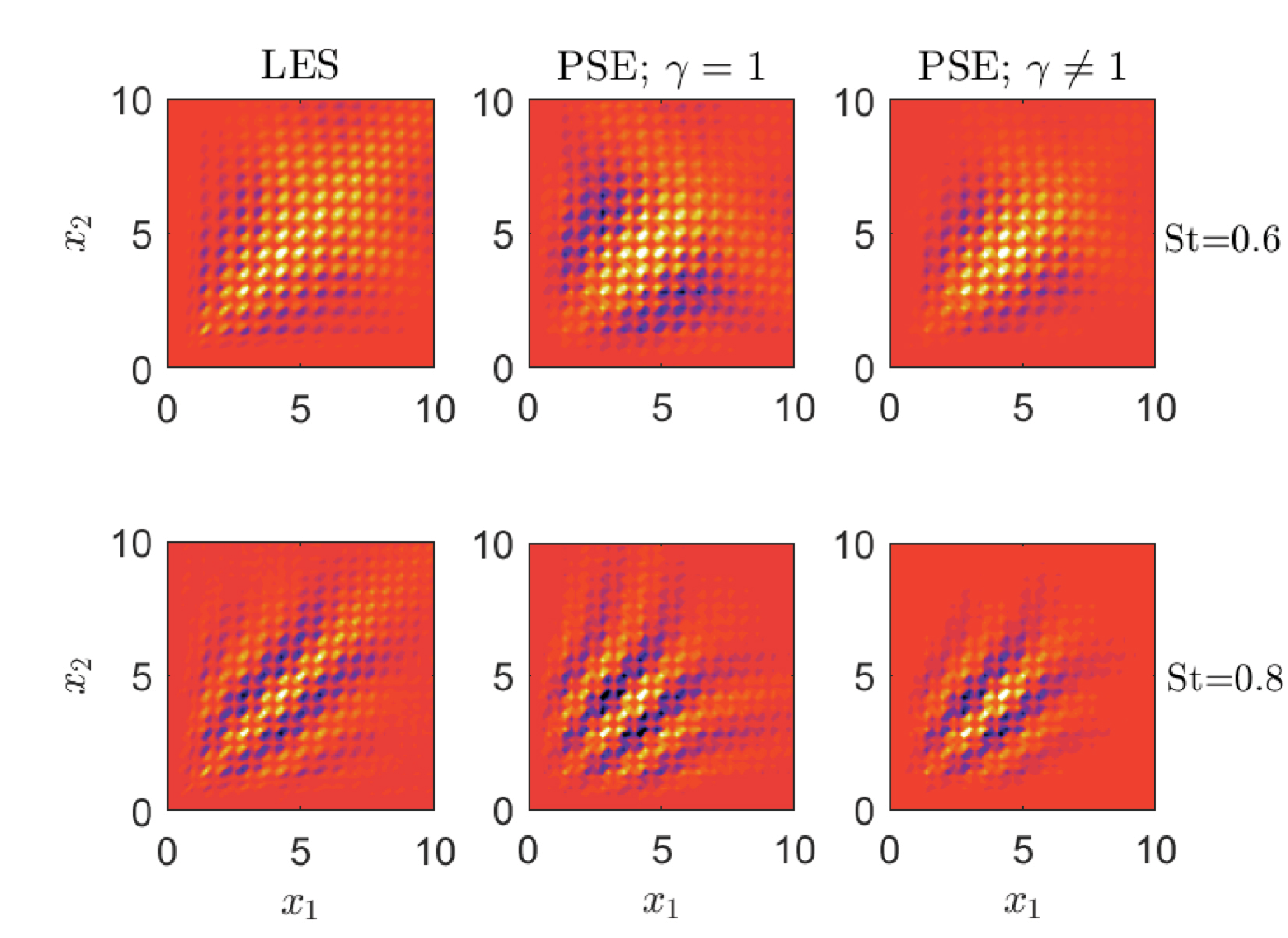}
    \caption{The normalised real components of the CSD of $S_{11}$ for $m=0$ and frequencies $St=0.6$ (top row) and $St=0.8$ (bottom row). The different reconstructed source models are LES (left column), PSE without coherence decay (centre column) and PSE with coherence decay (right column). Contours levels are from -0.5 to 0.5 and normalised by the maximum value.}
    \label{fig:LEScsd}
\end{figure}

We also note that the effect of coherence decay is not apparent in figure~\ref{fig:sourcemap}, even though it has significant effect on the far-field sound. To observe the effect of coherence decay, we present radially-integrated source CSDs as defined by equation~(\ref{eq:sij}), which are equivalent to a line-source approximation \citep{maia2019two, Wong2019}. Amplitudes are normalised for qualitative comparisons. The radially-integrated CSD of the LES source is shown in the left column in figure~\ref{fig:LEScsd}. The freckled appearance is consistent with the CSD of the nearfield pressure of a shock-containing jet \citep{suzuki2016wave, Wong2019}. Discrete peaks are present as the wavepacket interacts with the periodic shock-cell structure. A perfectly-coherent source (centre column) results in a spatially-broader CSD since the wavepacket is coherent over larger lengthscales. When coherence decay (right column) is incorporated into the source description, it narrows the CSD as expected \citep{cavalieri2014coherence, Wong2019}. The effect of coherence decay is to make the perfectly-coherent CSD more compact, and hence more similar to the LES model.

\section{Summary and conclusions} \label{sec:conclusion}
We present a two-point model for investigating the sound-source mechanisms of broadband shock-associated noise. Using the same interpretation as \citet{Tam1982}, BBSAN is assumed to be produced by the non-linear interaction of shocks and jet turbulence. We use a modified Pack and Prandtl vortex-sheet model to represent the quasi-periodic shock-cell structure. The turbulent component is modelled by a wavepacket, which is extracted directly from an LES database. To highlight the links to the underlying physical mechanisms, PSE solutions are also used to describe the statistical wavepacket shape. Both the P-P and PSE models are modified to match experimental and numerical flow data. Unlike previous models for BBSAN, the source parameters are solely determined by the turbulent flow field of the shock-containing jet. Acoustic measurements are not used to calibrate or alter the source.  

Two major conclusions may be drawn from the results of \S~\ref{sec:results}. Firstly, we have shown that a reduced-order representation of the equivalent source can provide largely accurate far-field predictions for BBSAN. This applies over a wide directivity range. Provided that shock-cell and mean flow profiles are available, only a single empirical constant is required to adjust the free amplitude of the linear PSE solutions. The efficacy of the simpler PSE-based approach is corroborated by agreement with the sound field features of the more complex, but complete, model using the LES CSD ($\pm$ 2dB/St at peak frequency). Examination of the results is aided by the availability of azimuthally-decomposed acoustic data. The encouraging comparisons between measurements and model predictions further support the BBSAN generation mechanism proposed by \citet{Tam1982}. %

Secondly, the results also provide some answers to the shortfalls of previous BBSAN models. As predicted by the line-source model of \citet{Wong2019}, the inclusion of the effects of wavepacket jitter and higher shock-cell modes is integral to predictive ability at higher frequencies and regions between the BBSAN peaks. We demonstrate the importance of these effects by directly quantifying and incorporating them into the description of the equivalent source. It seems clear that the `missing sound' observed at high frequencies by both \citet{ray2007sound} and \citet{suzuki2016wave} is due to the absence of higher shock-cell modes. The `dips' observed in the spectra of \citet{Tam1987} are removed following the inclusion of coherence decay. From a practical perspective, while inter-peak prediction is improved, accurate description of coherence profiles, decomposed in frequency and azimuth, is usually difficult to obtain. 

The artificial separation of the source into turbulent and shock components, however, means the effects of their interaction cannot be accounted for. Compelled by the modelling framework, $\bm{q}_{t}$ and $\bm{q}_{s}$ were both educed from separate ideally-expanded and shock-containing jets respectively. This may contribute to why, even with exact coherence information, the BBSAN predictions between the first and second peak at upstream angles underpredict the measured data. As hypothesised by \citet{Tam1987}, the discrepancy may be due to the inability for the model to capture shock-cell unsteadiness further downstream. This interaction between the two components should be investigated in future work. 

\section*{Acknowledgements}
The authors would like to thank Dr. Guillaume Brès at Cascade Technologies for providing the simulation database. The LES work was supported by ONR, with computational resources provided by DoD HPCMP. The research benefited from the Multi-modal Australian ScienceS Imaging and Visualisation Environment (MASSIVE) HPC facility, provided through the National Computational Merit Allocation Scheme. M.H.W., R.K. and D.E.M received funding from the Australian Research Council through the Discovery Projects scheme. M.H.W. is further supported by an Australian Government Research Training Program (RTP) Scholarship and the Endeavour Research Leadership Award – an Australian Government initiative. 

\section*{Declaration of interests}
The authors report no conflict of interest.

\appendix 
\section{Approximation of $T_{ij}$ for BBSAN} \label{app:A}
The substitution of the decomposed flow variables into $T_{ij}$ (equation~(\ref{eq:subTij})) is re-written below

\begin{equation}
    T_{ij} = (\bar{\rho}+\rho_s+\rho_t)( \bar{u}_i+u_{i,t}+u_{i,s})( \bar{u}_j+u_{j,t}+u_{j,s}) .
\end{equation}
By expanding out the terms we obtain,
\begin{eqnarray*} 
T_{ij} &=& \\
&&\bar{\rho}\bar{u}_i\bar{u}_j+\bar{\rho}\bar{u}_i u_{j,t}+\bar{\rho}\bar{u}_i u_{j,s}+\bar{\rho}\bar{u}_j u_{i,t}+ \\
&&\bar{\rho}u_{i,t} u_{j,t}+\bar{\rho}u_{i,t} u_{j,s}+\bar{\rho}\bar{u}_j u_{i,s}+\bar{\rho} u_{i,s} u_{j,t}+\bar{\rho} u_{i,s}u_{j,s} + \\
&&\rho_s\bar{u}_i\bar{u}_j+\rho_s\bar{u}_i u_{j,t}+\rho_s\bar{u}_iu_{j,s}+\rho_s\bar{u}_j u_{i,t}+ \\
&&\rho_s u_{i,t} u_{j,t}+\rho_s u_{i,t} u_{j,s}+\rho_s\bar{u}_j u_{i,s}+\rho_s u_{i,s} u_{j,t}+\rho_s u_{i,s} u_{j,s} + \\
&&\rho_t\bar{u}_i\bar{u}_j+\rho_t\bar{u}_i u_{j,t}+\rho_t\bar{u}_iu_{j,s}+\rho_t\bar{u}_j u_{i,t}+ \\
&&\rho_t u_{i,t}u_{j,t}+\rho_t u_{i,t}u_{j,s}+\rho_t\bar{u}_j u_{i,s}+\rho_t u_{i,s}u_{j,t}+\rho_t u_{i,s}u_{j,s}.
\end{eqnarray*}
To proceed, only the leading-order fluctuation terms are retained and higher-order ones are discarded. Furthermore, we only retain the interaction terms between turbulence and shocks (as these contribute to BBSAN). Thus, we can simplify the above expression such that

\begin{eqnarray} \label{eq:expandstress}
    T_{ij} \approx &\bar{\rho}(u_{i,t}u_{j,s}+u_{i,s}u_{j,t})+\bar{\rho_s}(\bar{u}_iu_{j,t}+\bar{u}_ju_{i,t})+\bar{\rho_t}(\bar{u}_iu_{j,s}+\bar{u}_ju_{i,s})+ \nonumber \\
    &\underbrace{ \left \{ \bar{\rho}\bar{u}_i\bar{u}_j+\bar{u}_i\bar{u}_j\rho_s + \rho_s(\bar{u}_iu_{j,s}+\bar{u}_ju_{i,s}) \right \}} \nonumber \\
\end{eqnarray}
The terms in the under-brace in equation~(\ref{eq:expandstress}) can be ignored because they are non-fluctuating and hence by definition cannot generate noise. We hence arrive at the approximated expression for the BBSAN stress tensor term
\begin{equation}
    T_{ij} \approx \bar{\rho}(u_{i,t}u_{j,s}+u_{i,s}u_{j,t})+\rho_s(\bar{u}_iu_{j,t}+\bar{u}_ju_{i,t})+\rho_t(\bar{u}_iu_{j,s}+\bar{u}_ju_{i,s}) .
\end{equation}

\section{Discussion on jet database parameters} \label{sec:note}
The effects on model predictions due to the variations between the databases is discussed in this appendix. Discrepancies, summarised in table~\ref{tab3}, include exit velocity, operating temperature (isothermal in LES, cold in experiments), Reynolds number and nozzle geometry. We again note that the LES and PIV flow fields are only used to inform the modelling choices in order to predict far-field BBSAN SPLs. No acoustic information is directly obtained or used from either of these databases.

\begin{figure}
    \centering
    \begin{subfigure}{0.45\textwidth}
    \includegraphics[width=1.0\linewidth]{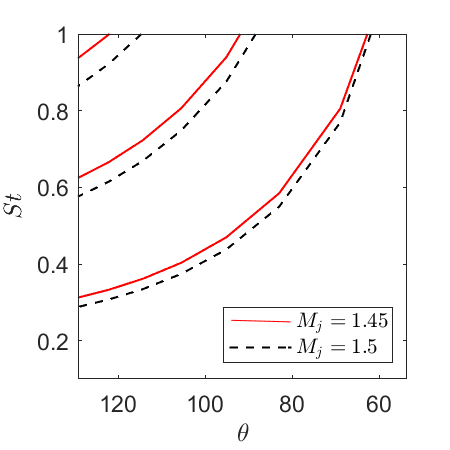}
    \caption{Variation in $M_j$.}
    \label{fig:Mj}
    \end{subfigure}
        \begin{subfigure}{0.45\textwidth}
    \includegraphics[width=1.0\linewidth,]{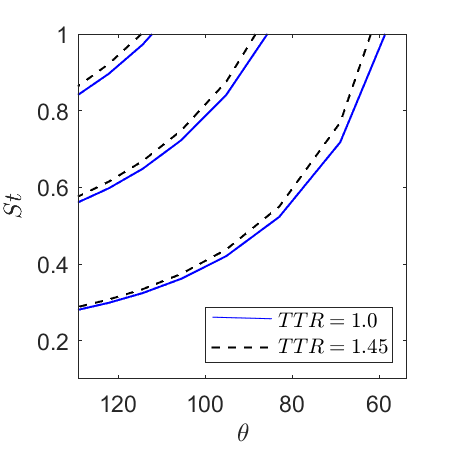}
    \caption{Variation in $TTR$.}
    \label{fig:Tj}
    \end{subfigure}
    \caption{Variation in peak BBSAN predictions as predicted by equation~(\ref{eq:HBFtam}) due to differences in jet parameters for the first three shock-cell modes ($n_s=1,2,3$). For the cold jet, a convection velocity of $u_c=0.7U_j$ was used for both plots while the relationship in equation~(\ref{eq:hot}) was used for the heated case.}
\end{figure}

As discussed in \S~\ref{sec:math}, the non-shock-containing components ($\bm{\bar{q}}$ and $\bm{\bar{q}_t}$) of the shock-containing jet should be obtained from the ideally-expanded case at the same $M_j$. To show the effect of using different values of $M_j$ on frequency, the non-dimensional form of equation~(\ref{eq:HBF}) is

\begin{equation}
    St_p = \frac{u_cD_j}{U_j} \left( \frac{1}{L_s(1-u_c/u_j\cos\theta)} \right),~ L_s \approx 1.3\beta ,
\end{equation}
where peak frequency is given by Strouhal $St_p$. Assuming a constant convection velocity of $u_c = 0.7U_j$, the only variable controlling the peak is the shock spacing $L_s$, which is approximately proportional to the off-design parameter $\beta$. The shock spacing of a $M_j=1.45$ jet is approximately 5\% shorter than that for $M_j=1.5$. The variation in the peak prediction is shown in figure~\ref{fig:Mj}, where we observe only a slight difference for the primary peak. For BBSAN intensity, which scales with $\beta^4$ \citep{harper1973noise}, the mismatch in $M_j$ results in a 1-2dB/St difference in sound pressure level.

The effect of temperature on BBSAN generation has previously been investigated in models \citep{Tam1990} and experiments \citep{kuo2015effects}. With relevance to peak frequency prediction in (\ref{eq:HBF}), heated jets have lower convection velocities and a shorter potential core. Despite these differences, the measurements of \citet{kuo2015effects} for underexpanded jets show either no change or a only a slight increase in peak frequency. This minor change is supported by the $St-\theta$ plot in figure~\ref{fig:Tj}. The convection velocity, as a function of temperature, is taken to be \citep{Tam1990}

\begin{equation} \label{eq:hot}
    u_c/U_j = 0.7-0.025(TTR-1), 
\end{equation}
where TTR represents the total temperature ratio, which is equal to 1.45 for the isothermal case and unity for a cold jet. This observation is corroborated by the measurements of \citet{wishart1995structure} who also found that the effect of varying temperature on shock structure is relatively small.

We note that all three databases are of fully-turbulent jets with $Re>400,000$, which \citet{viswanathan2002aeroacoustics} deems an appropriate threshold to avoid Reynolds number effects on the radiated sound field. Previous studies have also shown $Re$ having minimal effect on shock spacing and wavelengths \citep{tam1985multiple}. Similarly, \citet{hu1990flow} found that the evolution of large-scale structures at $Re=8000$ is similar to those in underexpanded jets at high Reynolds number. These observations give us some confidence that BBSAN may be considered independent of $Re$ for the databases investigated here.

Lastly, experimental studies have shown that nozzle geometry can strongly affect screech and resonant characteristics of a supersonic jet \citep{edgington2019aeroacoustic}. Screech is known to significantly influence the decay of the shock-cell structure and hence affects the production of BBSAN \citep{andre2013broadband}. Since both the acoustic and PIV databases use nozzles without screech suppression features, the intensity and frequency of BBSAN peaks are likely affected by the presence of screech. When interpreting the predictions of \S~\ref{sec:results}, it must be noted that the model does not account for such effects, which will remain a source of error.

While there remain tangible differences across the three databases, our goal is not to match predictions with a particular experiment, but rather to identify the underlying sound source mechanisms. Despite the minor mismatches, the results confirm the suitability of using these databases to inform our flow modelling choices. 

\bibliographystyle{jfm}

\bibliography{references}

\end{document}